\documentclass[twocolumn,twocolappendix]{aastex631}
\usepackage{natbib}
\usepackage{amsmath}
\usepackage{algorithm}
\usepackage{amsfonts}
\usepackage{mathrsfs}
\usepackage{amssymb}
\usepackage{color}
\usepackage{graphicx} 
\let\tablenum\relax
\usepackage{comment}

\newcommand{\lc}{\textcolor{black}}

\newcommand{\cha}{\textit{Chandra}}
\newcommand{\XMM}{{XMM-{\it{Newton}}}}
\newcommand{\NuSTAR}{{NuSTAR}}
\newcommand{\XSPEC}{{\tt{XSPEC}}}

\def \UXClumpy {{\tt \UXClumpy}}

\begin{document}

\title{PACHA: Probing AGN Coronae with High-redshift AGN}

\author[0000-0002-7791-3671]{Xiurui Zhao}
\affiliation{Cahill Center for Astrophysics, California Institute of Technology, 1216 East California Boulevard, Pasadena, CA 91125, USA}

\author[0000-0002-0273-218X]{Elias Kammoun}
\affiliation{Cahill Center for Astrophysics, California Institute of Technology, 1216 East California Boulevard, Pasadena, CA 91125, USA}

\author[0000-0002-6584-1703]{Marco Ajello}
\affiliation{Department of Physics and Astronomy, Clemson University, Kinard Lab of Physics, Clemson, SC 29634, USA}

\author[0000-0002-2624-3399]{Yanfei Jiang}
\affiliation{Center for Computational Astrophysics, Flatiron Institute, New York, NY 10010, USA}

\author[0000-0001-9094-0984]{Giorgio Lanzuisi}
\affiliation{INAF, Osservatorio di Astrofisica e Scienza dello Spazio di Bologna, via P. Gobetti 93/3, 40129 Bologna, Italy}

\author[0009-0000-9468-7277]{Anne Lohfink}
\affiliation{Department of Physics, Montana State University, P.O. Box 173840, Bozeman, MT 59717-3840, USA}

\author[0000-0002-2203-7889]{Stefano Marchesi}
\affiliation{Dipartimento di Fisica e Astronomia (DIFA), Università di Bologna, via P. Gobetti 93/2, 40129 Bologna, Italy}
\affiliation{Department of Physics and Astronomy, Clemson University, Kinard Lab of Physics, Clemson, SC 29634, USA}
\affiliation{INAF, Osservatorio di Astrofisica e Scienza dello Spazio di Bologna, via P. Gobetti 93/3, 40129 Bologna, Italy}

\author[0000-0001-5487-2830]{Elena Bertola}
\affiliation{INAF--OAA, Osservatorio Astrofisico di Arcetri, largo E. Fermi 5, 50127, Firenze, Italy}

\author[0000-0001-9379-4716]{Peter G. Boorman}
\affiliation{Cahill Center for Astrophysics, California Institute of Technology, 1216 East California Boulevard, Pasadena, CA 91125, USA}

\author[0000-0002-2115-1137]{Francesca Civano}
\affiliation{NASA Goddard Space Flight Center, Greenbelt, MD 20771, USA}

\author[0000-0001-8822-8031]{Luca Comisso}
\affiliation{Department of Astronomy and Columbia Astrophysics Laboratory, Columbia University, New York, NY 10027, USA}

\author[0000-0001-9604-2325]{Paolo Coppi}
\affiliation{Department of Astronomy, Yale University, New Haven, CT 06511, USA}

\author[0000-0003-2287-0325]{Isaiah S. Cox}
\affiliation{Department of Physics and Astronomy, Clemson University, Kinard Lab of Physics, Clemson, SC 29634, USA}

\author[0000-0001-5060-1398]{Martin Elvis}
\affiliation{Center for Astrophysics $|$ Harvard \& Smithsonian, 60 Garden Street, Cambridge, MA 02138, USA}

\author[0000-0001-8121-6177]{Roberto Gilli}
\affiliation{INAF, Osservatorio di Astrofisica e Scienza dello Spazio di Bologna, via P. Gobetti 93/3, 40129 Bologna, Italy}

\author[0000-0002-4226-8959]{Fiona A. Harrison}
\affiliation{Cahill Center for Astrophysics, California Institute of Technology, 1216 East California Boulevard, Pasadena, CA 91125, USA}

\author[0000-0001-6564-0517]{Ross Silver}
\affiliation{NASA Goddard Space Flight Center, Greenbelt, MD 20771, USA}

\author[0000-0003-2686-9241]{Daniel Stern}
\affiliation{Jet Propulsion Laboratory, California Institute of Technology, 4800 Oak Grove Drive, Pasadena, CA 91109, USA}

\author[0000-0003-3638-8943]{Nuria Torres-Alb\`a}
\affiliation{Department of Physics and Astronomy, Clemson University, Kinard Lab of Physics, Clemson, SC 29634, USA}

\author[0000-0002-6893-3742]{Qian Yang}
\affiliation{Center for Astrophysics $|$ Harvard \& Smithsonian, 60 Garden Street, Cambridge, MA 02138, USA}

\author[0000-0003-0232-0879]{Lizhong Zhang}
\affiliation{Center for Computational Astrophysics, Flatiron Institute, New York, NY 10010, USA}
\affiliation{School of Natural Sciences, Institute for Advanced Study, Princeton, NJ 08540, USA}

\begin{abstract}
The X-ray emission of active galactic nuclei (AGN) is generally attributed to the inverse Compton scattering in a compact corona. In local AGN, directly constraining coronal properties is challenging because the key spectral signature, the high-energy cutoff, is often outside the bandpass of even the most sensitive hard X-ray observatory, NuSTAR. High-redshift luminous quasars enable systematic constraints on the high-energy cutoff, as cosmological redshift shifts the spectral cutoff into the observable X-ray band. We present the first paper of our ``Probing AGN Coronae with High-redshift AGN" (PACHA) project based on quasi-simultaneous \NuSTAR\ and \XMM\ observations of 13 radio-quiet AGN at $z>1$. We constrain the high-energy cutoff and coronal temperature at 90\% confidence level for 10 and 9 sources, respectively. The sample exhibits a mean cutoff energy of $E_{\rm cut}$ = 101$\pm$21 keV and a mean coronal temperature of $kT_{\rm e}$ = 21.4$\pm$2.9 keV, both significantly lower than those measured in local AGN, while a higher mean optical depth ($\tau=4.8\pm0.2$). The uncertainties are at 1 $\sigma$. Within a hybrid coronal framework, such results indicate substantial efficient radiation cooling and/or non-thermal electron fraction. The low coronal temperatures observed in high-luminosity AGN are broadly consistent with radiation magnetohydrodynamic (MHD) simulations assuming purely thermal electrons, suggesting that non-thermal electrons are not the primary drivers of their coronal properties. Combining our high-redshift sample with local AGN, we find a potential anti-correlation between cutoff energy and both X-ray luminosity and black hole mass, with no significant dependence on Eddington ratio. 
\end{abstract}

\keywords{Active galactic nuclei, X-rays, Corona}

\section{Introduction} \label{sec:intro}
 The primary X-ray emission observed in active galactic nuclei (AGN) is believed to originate from the inverse-Compton scattering of optical/UV photons emitted from the accretion disk \citep[e.g.,][]{Haardt1991}. These photons are scattered by the so-called corona \citep{Vaiana1978,Haardt1993,Merloni2003}, a region surrounding the supermassive black hole (SMBH) that contains hot electrons with temperatures of about $10^8-10^9$~K (i.e., $\sim10-100$~keV). Based on studies of gravitationally lensed quasars \citep[e.g.,][]{Chartas2016} and X-ray eclipses \citep[e.g.,][]{Risaliti2007}, coronae are inferred to be very compact: with sizes on the order of $R_c\sim$10$^{-3}$--10$^{-5}$~pc or 10--100 $r_g$, where $r_g$ is the gravitational radius of the SMBH ($r_g$ = GM$_{\rm BH}$/$c^2$, M$_{\rm BH}$ is the black hole mass). Such compact scales make direct observation of the corona and its properties exceedingly challenging. 

In AGN coronae, high-energy photons upscattered by hot electrons can interact to produce electron-positron pairs when their center-of-mass energy exceeds $\sim$1\,MeV. The pair density increases with both the electron temperature ($kT_e$) and the coronal compactness ($\ell\propto L_c/R_c$, where $L_c$ is the coronal luminosity). When $kT_e$ or $\ell$ rises above a critical line in the $kT_e-\ell$ plane, the production of electron-positron pairs increases, leading to a reduction in the average electron temperature, and thus a reduction in electron-positron pair production \citep{Fabian2017}. Therefore, in the self-regulated coronal scenario, the electron temperature and compactness are expected to remain near the critical line predicted by the pair-production model. The critical line depends on both the ratio between thermal and non-thermal electrons in the corona and the ratio between the total coronal heating power and the input soft-photon power from accretion \citep{Fabian2017}. Therefore, a sample of AGN with well constrained coronal properties provide a powerful means of systematically probing the physics of the corona and disk--corona coupling.

\begin{figure}[t] 
\centering
\includegraphics[width=.45\textwidth]{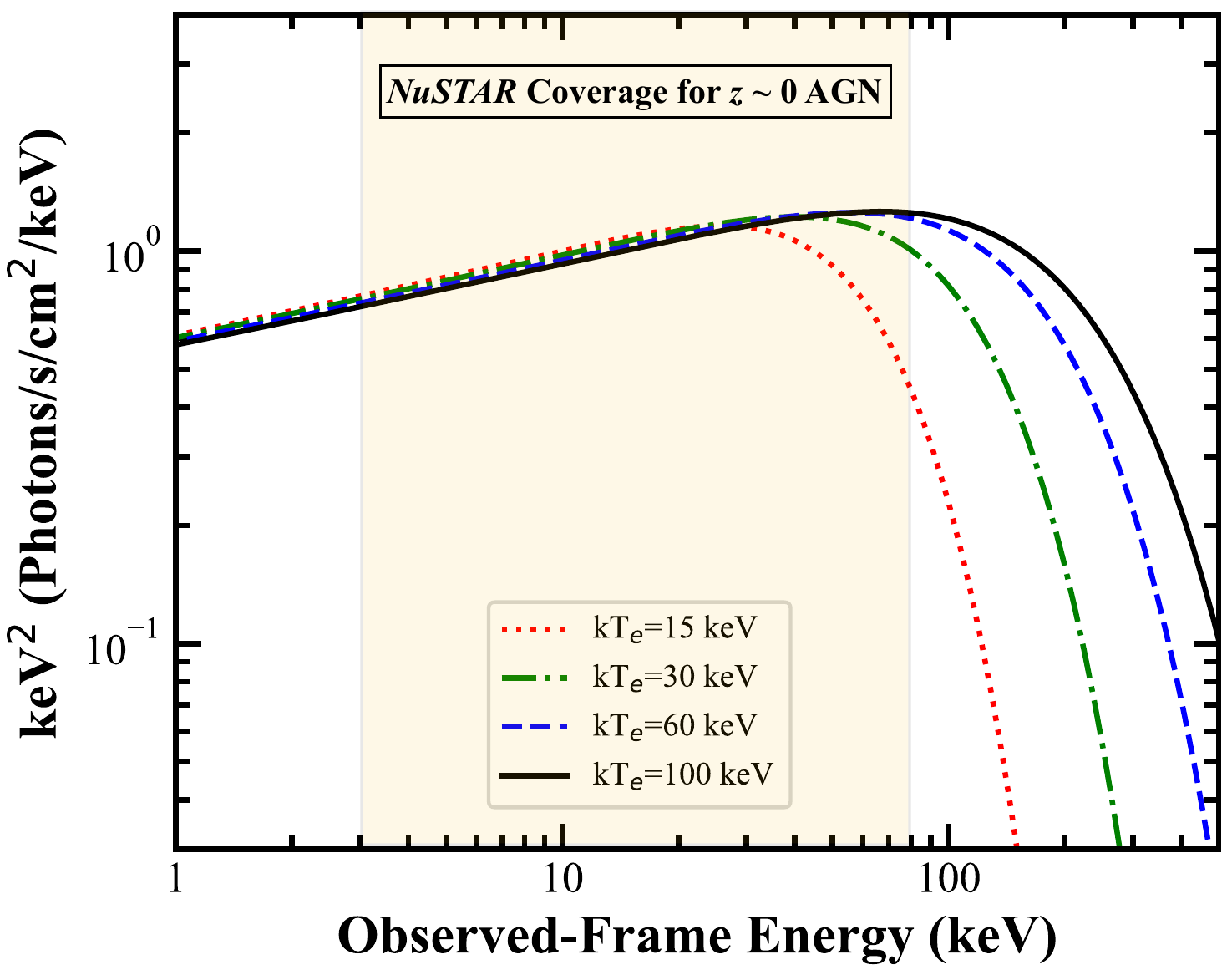}
\caption{AGN spectra computed assuming a Comptonization model with different coronal temperatures. For $z\sim0$ sources with high $kT_{\rm e}$, the high-energy cutoff is shifted above the \NuSTAR\ bandpass (3--78~keV; yellow shaded region), limiting our ability to constrain the coronal temperature.}
\label{fig:coverage}
\end{figure}

The X-ray spectrum of the corona is commonly modeled as a power law with an exponential cutoff ($E_{\rm cut}$) at tens to hundreds of keV. In thermal Comptonization models, this cutoff is related to the electron temperature, with $E_{\rm cut}\sim$2--3~$kT_e$ depending on the coronal optical depth $\tau$ \citep{Petrucci2001}. Therefore, the high-energy cutoff in the X-ray spectrum (typically at a few tens to hundred keV) can be directly linked to the properties of the corona. Thanks to the launch of \NuSTAR\ \citep{harrison}, whose hard X-ray ($\ge$10~keV) sensitivity is orders of magnitude greater than previous missions, a large sample of AGN can now be observed with unprecedented data quality in the hard X-ray band \citep[e.g.,][]{harrison15,Civano_2015,Lansbury_2017,Marchesi2018,Nuria,Greenwell2024}. The majority of sources observed to date are low-luminosity AGN ($L_{\rm 2-10\,keV}\approx$ 10$^{42}$--10$^{44}$ erg\,s$^{-1}$) in the nearby Universe ($z<0.1$). As a result, only lower limits to the coronal temperature were obtained for most targets \citep[e.g.,][]{Ricci2018,Tortosa2018,Balokovic2020,Kang_2022,Kamraj2022} due to the limited bandpass of \NuSTAR\ (3--78~keV, Fig.~\ref{fig:coverage}).

High-redshift quasars, which are predominantly high-luminosity sources ($L_{\rm 2-10\,keV}>$ 10$^{45}$~erg~s$^{-1}$), enable systematic measurements of coronal temperatures, optical depth, and thermal to non-thermal electron ratios thanks to the cosmological redshifting of the high-energy cutoff into the observable band. Therefore, high-redshift quasars constitute the best sample to constrain the AGN coronal properties and probe the electron population in coronae. So far, four bright radio-quiet quasars at $z>$1 have deep \NuSTAR\ observations. All four sources have well-constrained coronal temperatures \citep{Lanzuisi2016,Lanzuisi2019,Bertola2022} and their coronal temperatures are significantly below the critical threshold in the $kT_e-\ell$ plane predicted by a purely thermal corona model assuming a spherical coronal geometry. However, the small number of high-$z$ quasars with well-constrained coronal properties currently limits our ability to draw statistically robust conclusions about the nature of AGN coronae and their Comptonization processes.

In this work, we present the first paper in a series of PACHA\footnote{PACHA means ``world” or ``spacetime” in Inca cosmology.} project: Probing AGN Coronae with High-redshift AGN. The paper is organized as follows: in Section~\ref{sec:source_selection}, we present the sample selection criteria. In Section~\ref{sec:spec_ana}, we present the X-ray spectral analysis and results of the 13 sources in our sample. In Section~\ref{sec:discussion}, we compare the coronal properties of our high-$z$, high-luminosity sample with those of the local, low-luminosity AGN. We further investigate the electron distribution in the AGN coronae. In addition, we discuss the measured coronal properties \lc{within the framework of a radiation magnetohydrodynamic (MHD) model}. In addition, we discuss the impact of gravitational and relativistic effects on the coronal property measurements. We also probe the potential constraints that can be placed on the coronal height and size using the derived coronal properties. Finally, we compare the X-ray to bolometric luminosity relation derived in this sample with previous measurements.
All uncertainties quoted in this paper are at the 90\% confidence level unless otherwise stated. The adopted cosmological parameters are as follows: $H_0$ = 70 km s$^{-1}$ Mpc$^{-1}$, $\Omega_M$ = 0.3, and $\Omega_\Lambda$ = 0.7.

\section{Source Selection}\label{sec:source_selection}

Large \NuSTAR\ joint \XMM\ programs were awarded in both \NuSTAR\ cycle 8 and cycle 9 (PI: Zhao, proposal IDs 8170 and 9182) to observe a total of eight bright AGN at high redshifts ($z\ge1$) in order to measure their coronal properties. The eight targets were selected from the Sloan Digital Sky Survey (SDSS) Data Release 16 AGN catalog \citep{Wu2022}, which includes more than 750,000 type 1 AGN. Their X-ray fluxes were derived from the \cha\ source catalog Release 2.1 \citep[CSC 2.1,][]{Evans2024} and the fourth \XMM\ serendipitous survey \citep[4XMM-DR12,][]{Webb2020}. Only bright (2--10 keV flux F$_{\rm 2-10}\ge3$~$\times$~10$^{-13}$ erg\,cm$^{-2}$\,s$^{-1}$) sources were selected as their cutoff energies can be well constrained by \NuSTAR\ within a reasonable exposure ($<$200~ks). 

The hard X-ray emission of blazars is dominated by inverse Compton scattering emission from the jet rather than the hot corona \citep{Elvis1994,Padovani:2017we}. Therefore, we excluded quasars which are identified as blazars in the most recent Roma-BZCAT catalog \citep{Massaro2015}. We further excluded blazar candidates selected based on {\it WISE} mid-infrared colors \citep{Abrusco2019}. To ensure pure coronal emissions in hard X-rays, we require radio-quiet quasars with radio loudness\footnote{Radio loudness is defined as the ratio of the 5~GHz flux density to the 4400\AA\ flux density.} $RL<$10 for our sample.

In addition to the eight proposed targets, four $z>1$ radio-quiet AGN \citep{Lanzuisi2016,Lanzuisi2019,Bertola2022}, previously observed with \NuSTAR\ and \XMM\ during \NuSTAR\ Cycles 3 and 4 (PI: Lanzuisi; proposal IDs 3197 and 4139), satisfy our target-selection criteria and are therefore included in our sample. The X-ray spectra of these sources are re-analyzed in this work. Furthermore, we include SDSS~J145453.53+032456.8, a $z=2.375$ radio-quiet AGN with deep \NuSTAR\ and \XMM\ coverage obtained during the \NuSTAR\ prime mission, which has not been analyzed in any previous studies.

A total of 13 targets are included in the PACHA sample. Their source properties, including redshifts and black hole masses (M$_{\rm BH}$) are summarized in Table~\ref{Tab:source}. 
The black hole masses are mainly derived using the single-epoch spectroscopic method as reported in \citet{Wu2022}; exceptions include 2MASS J1614346+470420 and B1422+231, for which the black hole masses are adopted from \citet{Shen2011} and \citet{Assef2011}, respectively. In addition, the redshift and M$_{\rm BH}$ of WISEA J163428.99+703132.4 are measured in this work (Section~\ref{sec:opt_spec}). While the measurement uncertainties for single-epoch virial masses are typically small ($<$0.1 dex), the method is subject to systematic uncertainties of up to $\sim$0.4~dex \citep{Shen2013}. Consequently, we adopt a uniform uncertainty of 0.5~dex for black hole masses. The black hole mass of APM 08279+5255 was measured using the reverberation mapping method \citep{Saturni2016}, which is known to have much smaller uncertainties. 

We report the details of the \NuSTAR\ and \XMM\ observations of all 13 targets analyzed in this work in Table~\ref{Tab:obs}. The sample includes four gravitationally lensed AGN. The magnification factors of J0921+2854 ($\mu$ = 4), B1422+231 ($\mu$ = 20), and APM 08279+5255 ($\mu$ = 4) are adopted from \citet{Chartas2021}, \citet{Assef2011}, and \citet{Riechers2009}, respectively. The magnification factor of J0913+5259 ($\mu$ = 8.9) is estimated using the Singular Isothermal Ellipsoid (SIE) lens model \citep{Kormann1994} and data from archival \cha\ observation. The derived luminosity, Eddington ratio, and compactness require an accurate measurement of the magnification factor, whereas the cutoff energy, coronal temperature, photon index, and optical depth can be constrained independently of the magnification factor.

\begingroup
\renewcommand*{\arraystretch}{1.4}
\begin{table*}
\begin{center}
\scriptsize
\caption{Summary of Source Properties. The coronal properties of the first nine sources are studied for the first time in this work. The last four sources have been studied in previous works \citep{Lanzuisi2016,Lanzuisi2019,Bertola2022}, but we re-analyze their spectra following the method used in this work. The gravitationally lensed targets are labeled with $^G$.
Black hole mass is in unit of Solar mass.
F$_{2-10}$ is the 2--10 keV flux in 10$^{-13}$~erg~s$^{-1}$~cm$^{-2}$ derived from M CP.
$\Gamma$, $E_{\rm cut}$, and $R$ are the photon index, high-energy cutoff (in keV), and reflection scaling factor derived from M PEX, respectively.
$kT_{\rm e}$ and $\tau$ are the electron temperature and optical depth of coronae in keV derived from model CP, respectively.
L$_{\rm 2-10}$ is the 2--10~keV intrinsic (delensed) luminosity in $10^{45}$ erg s$^{-1}$ derived from model CP.
L$_{\rm bol}$ is the bolometric intrinsic luminosity in $10^{45}$ erg s$^{-1}$ integrated from 1~$\mu$m to $\sim$0.12~\AA\ (100~keV).
$\ell$ is the coronal compactness. The uncertainties of the best-fit parameters were estimated using the {\tt error} command in XSPEC and correspond to the 90\% confidence level for one parameter of interest ($\Delta C$ = 2.71).}\label{Tab:source}
  \begin{tabular}{cccccccccccccc}
       \hline
       \hline
    Target&$z$&log(M$_{\rm BH}$)&F$_{2-10}$&$\Gamma$&$E_{\rm cut}$&$kT_{\rm e}$&$\tau$&$R$&L$_{\rm 2-10}$&L$_{\rm bol}$&$\lambda_{\rm Edd}$&$\ell$\\ 
    \hline
   J0948+4323&1.893&9.6$\pm$0.5&3.0$_{-0.1}^{+0.2}$&1.56$\pm$0.04&59$_{-18}^{+20}$&14$_{-3}^{+5}$&5.7$\pm$0.9&0.05$_{-u}^{+0.27}$&5.5$\pm$0.1&95&0.2$^{+0.3}_{-0.1}$&100$_{-68}^{+220}$\\
    J1225+2235&2.059&10.0$\pm$0.5&4.7$\pm$0.1&1.76$_{-0.08}^{+0.09}$&70$_{-23}^{+100}$&15$_{-4}^{+14}$&5$\pm$1&0.3$_{-u}^{+0.4}$&4.7$\pm$0.1&480&0.4$^{+0.8}_{-0.3}$&33$_{-23}^{+71}$\\
    J1634+7031&1.337&9.9$\pm$0.5&8.5$\pm$0.3&2.11$_{-0.10}^{+0.11}$&66$_{-18}^{+28}$&15$_{-3}^{+6}$&3.6$_{-0.6}^{+0.8}$&0.99$_{-0.44}^{+0.63}$&8.7$\pm$0.3&630&0.6$^{+1.3}_{-0.4}$&84$_{-57}^{+180}$\\
J1454+0324&2.375&9.6$\pm$0.5&5.3$_{-0.2}^{+0.1}$&1.81$\pm$0.03&140$_{-50}^{+160}$&27$_{-7}^{+36}$&2.9$_{-1.0}^{+1.0}$&0.02$_{-u}^{+0.16}$&20.6$_{-0.3}^{+0.4}$&180&0.2$^{+0.6}_{-0.1}$&390$_{-270}^{+850}$\\    
J0913+5259$^G$&1.377&9.3$\pm$0.5&16.2$_{-0.4}^{+0.3}$&1.72$_{-0.02}^{+0.03}$&42$_{-5}^{+7}$&11$\pm1$&5.6$\pm$0.3&0.40$_{-0.15}^{+0.17}$&$\sim$1.8&$\sim$70&$\sim$0.3&88$_{-60}^{+190}$\\
J0921+2854$^G$ &1.410&9.4$\pm$0.5&7.9$_{-0.2}^{+0.3}$&1.57$\pm$0.03&71$_{-16}^{+27}$&13$\pm2$&5.9$\pm$0.6&0.24$_{-0.14}^{+0.13}$&$\sim$1.8&$\sim$70&$\sim$0.2&50$_{-34}^{+110}$\\
J1458+5254&1.746&9.6$\pm$0.5&1.5$_{-0.1}^{+0.2}$&1.94$\pm$0.08&$>$65&$>$14&$<$5.0&0.1$_{-u}^{+0.5}$&3.0$\pm$0.1&100&0.2$^{+0.4}_{-0.1}$&58$_{-40}^{+130}$\\
J1233+1612&1.331&9.5$\pm$0.5&1.8$\pm$0.1&1.74$\pm$0.06&$>$49&$>$13&$<$7.5&0.56$_{-0.45}^{+0.60}$&1.44$\pm$0.05&11&0.02$^{+0.04}_{-0.01}$&39$_{-27}^{+84}$\\
J1259+3423&1.378&9.8$\pm$0.5&2.8$\pm$0.1&2.02$_{-0.05}^{+0.06}$&$>$70&$>$20&$<$4.2&1.7$_{-0.6}^{+0.8}$&2.4$\pm$0.1&150&0.2$^{+0.4}_{-0.1}$&30$_{-21}^{+65}$\\
    \hline
J1614+4704&1.865&9.8$\pm$0.5&4.5$\pm$0.1&1.94$\pm$0.03&97$_{-27}^{+57}$&21$_{-4}^{+16}$&3.3$_{-1.0}^{+0.6}$&1.0$\pm$0.3&8.1$\pm$0.2&390&0.5$^{+1.0}_{-0.3}$&96$_{-66}^{+210}$\\
B1422+231$^G$&3.629&9.7$\pm$0.5&6.4$\pm$0.2&1.57$\pm$0.06&58$_{-9}^{+10}$&16$\pm$2&5.2$\pm$0.5&0.44$_{-0.17}^{+0.21}$&$\sim$2.2&$\sim$91&$\sim$0.1&33$_{-23}^{+71}$\\
APM 08279$^G$&3.912&10.0$\pm$0.1&2.6$\pm$0.1&2.15$_{-0.32}^{+u}$&98$_{-46}^{+450}$&20$_{-3}^{+23}$&3.3$_{-0.9}^{+0.4}$&3.1$_{-1.4}^{+3.3}$&$\sim$5.0&$\sim$280&$\sim$0.2&37$_{-8}^{+10}$\\
PG 1247+267&2.048&9.8$\pm$0.5&2.8$_{-0.2}^{+0.1}$&2.29$_{-0.08}^{+0.09}$&81$_{-28}^{+83}$&$>$19&$<$3.5&2.1$_{-0.9}^{+1.3}$&7.6$\pm$0.7&1000&1.3$_{-0.9}^{+2.7}$&100$_{-70}^{+220}$\\
           \hline
              \hline
\end{tabular}
\end{center}
\end{table*}
\endgroup

\section{Spectral Analysis \& Results} \label{sec:spec_ana}
\subsection{Spectral Analysis}
The data and X-ray spectra were reduced and extracted following the standard pipelines, which are presented in detail in \ref{sec:DR}.
The spectra were fitted with both phenomenological and physical Comptonization models. We used \XSPEC\ \citep{Arnaud1996} v12.13.1 to fit the \NuSTAR\ and \XMM\ spectra simultaneously, using the $C$-statistic \citep{Cash1979} to find the best-fit values for each model. The setup of the models is described in the following subsections. The \NuSTAR\ and \XMM\ spectra were grouped with the optimal binning scheme developed in \citet{Kaastra2016} using the {\tt ftgrouppha} tool.

\subsubsection{Phenomenological Models}
We started from phenomenological models, assuming that the intrinsic emission of AGN can be characterized as a power law ({\tt zpowerlw}). This model (hereafter, ``M powerlw'') in \XSPEC\ nomenclature is therefore:
%
%
\lc{\begin{equation}
\texttt{M powerlw} = \texttt{const} \times \texttt{TBabs} \times \texttt{zpowerlw}
\end{equation}}

where {\tt const} represents the cross calibration between \NuSTAR\ and \XMM\ modeled by a constant and is denoted as $C_{\rm N/X}$. {\tt TBabs} models the Galactic absorption in the line-of-sight to the source, where the column density is obtained using the {\tt nh} task \citep{nh} in \texttt{HEAsoft}. 

J1634+70313 presents an excess in soft X-rays ($\le$1~keV), which was modeled with a blackbody {\tt zbbody} model and is further discussed in Section~\ref{sec:spectral_analysis}. We tested for intrinsic absorption and found that only one source, APM~08279+5255, has a constrained intrinsic column density, which is modeled using the {\tt zTBabs} component.

We then fit the spectra assuming that the intrinsic power law emission has a high-energy cutoff ({\tt zcutoffpl}). The cutoff energy of the spectra can be used to infer the temperature of the corona. The model (hereafter, ``M cutoff'') in \XSPEC\ nomenclature is:
%
\lc{\begin{equation}
\label{eq:cutoff}
\texttt{M cutoff} = \texttt{const} \times \texttt{TBabs} \times \texttt{zcutoffpl}
\end{equation}}
Reflection from the accretion disk and/or the torus is found in X-ray spectra of both type 1 and type 2 AGN \citep[e.g.,][]{Ricci2017}. At high redshift, reflection contributions must be carefully taken into account, as the prominent Compton hump feature (rest-frame $\sim$10--30 keV) is redshifted into the spectral range covered by \XMM\ and \NuSTAR. In this work, we employed {\tt pexmon} \citep{Nandra2007} following \citet{Ricci2017} and \citet{Akylas2021}, which describes the coronal intrinsic emission (modeled by an exponentially cutoff power law) reflected from neutral material. {\tt pexmon} combines the continuum from {\tt pexrav} \citep{pexrav} with self-consistently generated emission lines from iron and nickel. 

We fixed the inclination angle to a face-on orientation ($incl = 0^\circ$), appropriate for type 1 AGN, to reduce the degeneracy among different parameters. We found that the choice of alternative inclination values (e.g., $incl$ = 19$^\circ$) has minimal impact on the best-fit values of the other model parameters. The iron abundance of the accretion disk was fixed to the Solar value (Fe$_{\rm abund}$ = 1)as it cannot be constrained in most cases. The only exception is J0921+2854, for which a supersolar iron abundance is required, as discussed in the \ref{sec:spectral_analysis}. The reflection scaling factor $R$ is set to be positive ($R>0$) to include both the intrinsic continuum and reflection. The model (hereafter, ``M PEX'') in \XSPEC\ nomenclature is:
%
%
\lc{\begin{equation}
\label{eq:pex}
\texttt{M PEX}=\texttt{const} \times \texttt{TBabs} \times \texttt{pexmon}
\end{equation}}

\subsubsection{Comptonization Model}
We fit the spectra with a physical Comptonization model using {\tt ThComp} \citep{Zdziarski2020}. {\tt ThComp} describes the comptonization of seed photons from the accretion disk \citep[{\tt diskbb},][]{Mitsuda1984} by thermal electrons emitted by a spherical source. We tested different values of the disk temperature, and only a marginal effect was found. Therefore, we fixed $T_{\rm disk}$ at 10~eV. We fix the covering fraction in the {\tt ThComp} model at $cov\_frac$ = 1, implying that all seed photons are Comptonized. 

The reflection component was modeled with {\tt xillverCp} \citep{Garcia2013}, which assumes a Comptonization incident spectrum. The density of the accretion disk was fixed at the default value of 10$^{15}$~cm$^{-3}$. Given the current data quality, allowing the disk density to vary has only a minimal effect on the best-fitting results. The coronal temperature of {\tt xillverCp} is linked to the coronal temperature of {\tt ThComp}. We fixed the inclination angle to a face-on orientation ($incl = 19^\circ$, the minimum allowed value in {\tt xillverCp}). The iron abundances were set following M PEX. {Since the reflection component contributes only marginally to the overall spectrum, the ionization parameter was fixed at $\log(\xi)=2$, unless allowing it to vary significantly improved the fit or the data quality was sufficient to constrain it independently.} Both these scenarios are further discussed in \ref{sec:spectral_analysis}. The model (hereafter, ``M CP'') in \XSPEC\ nomenclature is:
%
%
\lc{\begin{equation}
\label{eq:com}
\texttt{M CP}=\texttt{const} \times \texttt{TBabs} \times (\texttt{ThComp} \times \texttt{diskbb} + \texttt{xillverCP})
\end{equation}}
We computed the optical depth ($\tau$) of the corona following Eq.~14 in \citet{Zdziarski2020}, with 
\begin{equation}
\label{eq:tau}
\Gamma = \sqrt{\frac{9}{4} + \frac{1}{\bar{u}\,\theta\,\xi(\theta)}}-\frac{1}{2}
\end{equation}
where $\theta$ = $kT_{\rm e}$/511~keV, and $\xi(\theta)$ = 1+$\theta$+3$\theta^2$, $\bar{u}$ = $a_1\tau$ + $a_2\tau^2$, $a_1$ = 1.2/(1+$\theta$+5$\theta^2$), and $a_2$ = 0.25/$\xi(\theta)$. Note that this equation is calculated assuming a spherical geometry and a purely thermal corona. 

\begin{figure}[t] 
\centering
\includegraphics[width=.49\textwidth]{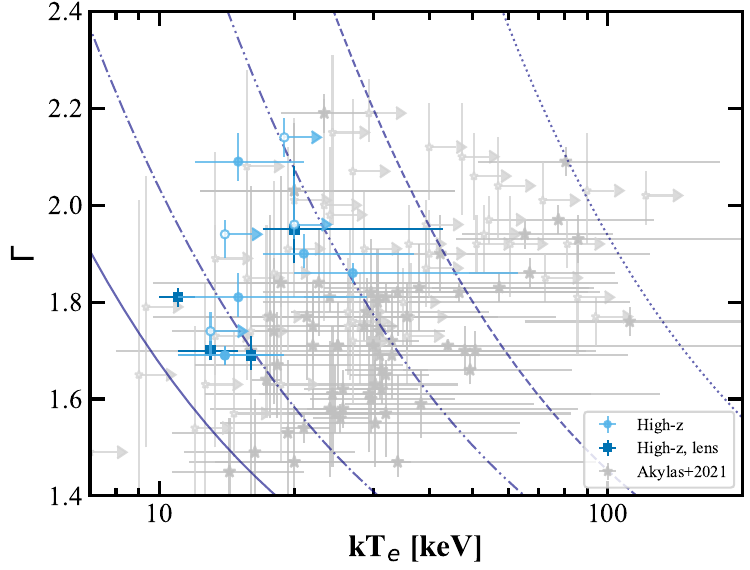}
\caption{Photon index as a function of the coronal temperature of the AGN coronae from our PACHA sample (blue) and the \citet{Akylas2021} sample (grey). The purple lines indicate different optical depths calculated using Eq.~\ref{eq:tau}.}
\label{fig:tau}
\end{figure}  

\subsection{Results}
We fitted the \NuSTAR\ and \XMM\ spectra of each source using both phenomenological and Comptonization models. The cutoff energy and coronal temperatures are well constrained in nine out of the 13 high-$z$ AGN, confirming that high-$z$, luminous AGN are excellent targets to constrain the coronal properties of AGN. For the remaining four targets, lower-limits on the coronal temperature are obtained, which is largely due to the flux of these targets being significantly lower than expected. The 2--10~keV fluxes of all four targets are below $3\times10^{-13}$~erg~cm$^{-2}$~s$^{-1}$. Based on extensive simulations, we find that this value corresponds to the approximate flux threshold required to reliably constrain coronal properties with $\sim$100--150~ks \NuSTAR\ exposures.
In \ref{sec:spectral_analysis}, we report the best-fit values of the spectra of each source using all four models (Table~\ref{Tab:all_source}) and we plot their best-fit spectra in Fig.~\ref{fig:spectrum1} to Fig.~\ref{fig:spectrum3}.

We summarize the best-fit high-energy cutoff ($E_{\rm cut}$), coronal temperature ($kT_{\rm e}$), inferred coronal optical depth ($\tau$), and 2--10~keV intrinsic luminosity for each source in Table~\ref{Tab:source}. {The confidence contours of the parameters $\Gamma$ vs. $E_{\rm cut}$, $R$ vs. $E_{\rm cut}$ for PEX model and $\Gamma$ vs. $kT_{\rm e}$ for CP model of the PACHA targets are plotted in Fig.~\ref{fig:contour1} to \ref{fig:contour3}.}

The reported coronal temperature, luminosity, and thus the derived compactness have not been corrected for general relativistic (GR) effects close to the SMBH, such as gravitational redshift and light bending \citep{Fabian2015}. These GR effects are further discussed in Section~\ref{sec:GR}.

\begin{figure}[t] 
\centering
\includegraphics[width=.49\textwidth]{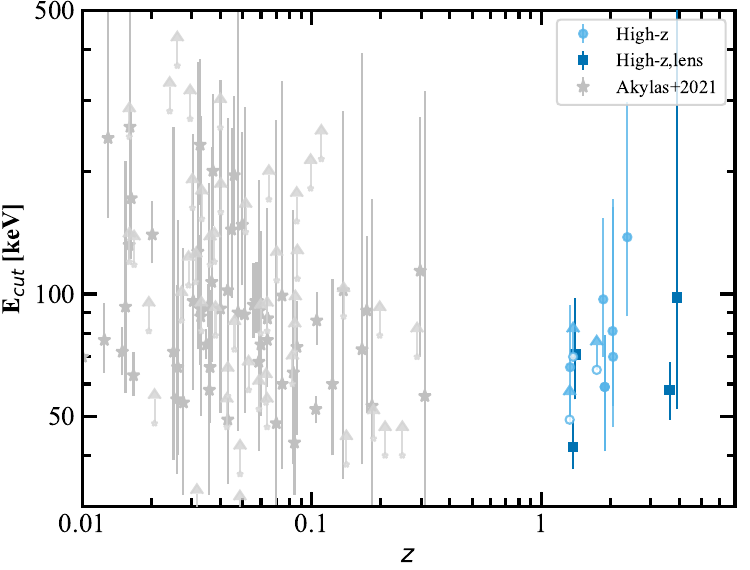}
\caption{Cutoff energy as a function of redshift for high-luminosity sources in our sample (blue) and local, low luminosity AGN (grey) adopted from \citet{Akylas2021}. The lower limits are calculated at the 90\% confidence level for both samples (open symbols).}
\label{fig:kTe_z}
\end{figure}   
\section{Discussion}\label{sec:discussion}
\subsection{Comparison with Local AGN}
We compared the measured properties of our high-$z$, luminous AGN with those measured in 117\footnote{The sample initially comprised 118 sources; however, the best-fit results were reported only for 117 sources (excluding ESO 344-G016) in \citet{Akylas2021}.} local, lower luminosity AGN selected from the {\it Swift}-BAT (14--195~keV) 70-month catalog \citep{Ricci2017}, measured using the \NuSTAR\ and soft X-ray observations in \citet{Akylas2021}.  

Approximately 50\% of the coronal cutoff energies in the local Seyfert galaxies are not constrained, with only a lower limit on the cutoff energy, probably due to the intrinsically high coronal temperature or cutoff energy. This highlights the effectiveness of using high-$z$, luminous AGN to probe coronal properties. We exclude NGC 4051 from the sample as its cutoff energy lower-limit ($E_{\rm cut}>846$~keV) is much higher than the \NuSTAR\ bandpass coverage and is also much higher than the mean value of the entire sample. We plot the cutoff energy as a function of the redshift of our high-$z$, luminous sample with the \citet{Akylas2021} sample in Fig.~\ref{fig:kTe_z}. 

\begin{figure}[t] 
\centering
\includegraphics[width=.475\textwidth]{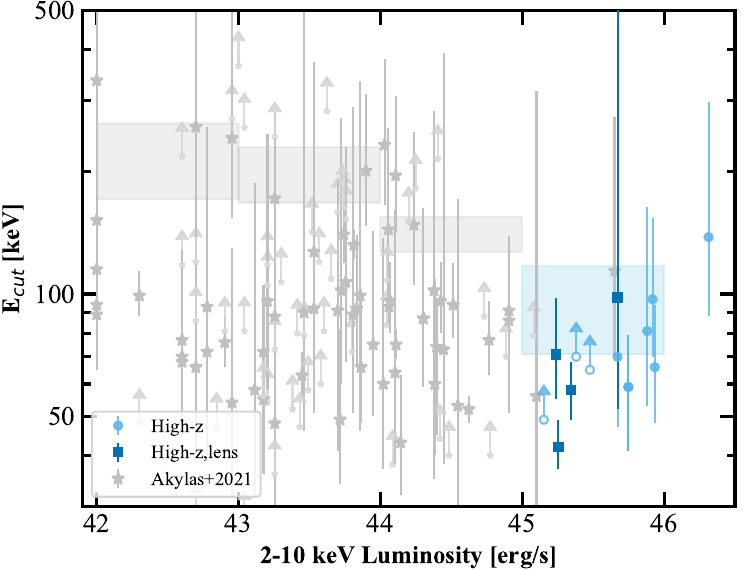}
\includegraphics[width=.475\textwidth]{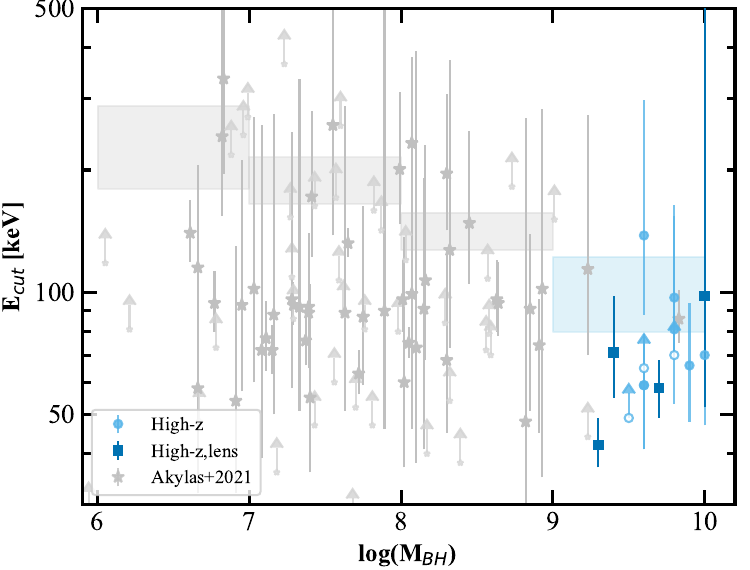}
\includegraphics[width=.475\textwidth]{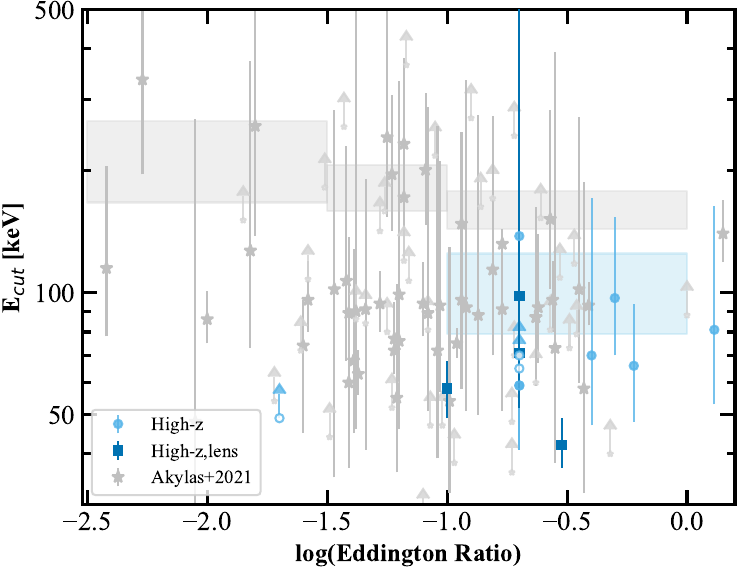}
\caption{Cutoff energy as a function of 2--10~keV luminosity (upper), black hole mass (middle), and Eddington ratio (bottom) of sources in our PACHA sample (blue) and {\it Swift}-BAT selected AGN (grey) from \citet{Akylas2021}. The means and their 1$\sigma$ uncertainties in different luminosity, Eddington ratio, and black hole mass bins are plotted in shaded regions.}
\label{fig:lum_Edd_cut}
\end{figure}

We plot in Fig.~\ref{fig:lum_Edd_cut} the cutoff energy as a function of the 2--10~keV intrinsic luminosity, for both the sources in our sample and the {\it Swift}-BAT local AGN from \citet{Akylas2021}. {As both the \citet{Akylas2021} sample and our PACHA sample contain censored measurements (lower limits on $E_{\rm cut}$), we estimate the mean cutoff energy using the Kaplan-Meier estimator implemented in the ASURV package \citep{Feigelson1985,Isobe1986}, following \citet{Akylas2021}. The Kaplan-Meier estimator is a non-parametric survival-analysis technique that incorporates both detections and censored measurements when estimating the underlying distribution. \citet{Akylas2021} further validated this approach through extensive Monte Carlo simulations, showing that the intrinsic cutoff-energy distribution inferred from the simulations is broadly consistent with the Kaplan-Meier estimate for their sample. 

The uncertainty on the Kaplan-Meier mean does not propagate the statistical uncertainties of the individual spectral fits. To propagate the statistical uncertainties of the individual measurements into the uncertainty of the sample mean, we performed a Monte Carlo analysis. For each source with a constrained coronal temperature (or cutoff energy), we generated 10,000 random realizations by sampling from the asymmetric 90\% confidence intervals, approximated as a split normal distribution. Sources with unconstrained values (lower limits) were kept fixed in each realization. For every realization, we recalculated the Kaplan--Meier mean using the ASURV package. The final sample mean was taken as the mean of the resulting distribution of Kaplan--Meier means, while the uncertainty was estimated from the standard deviation assuming an approximately Gaussian distribution. This procedure propagates the measurement uncertainties of individual sources while consistently accounting for the censored data through the Kaplan--Meier estimator.}

{The cutoff energies of both the PACHA sample and local AGN sample are taken from the {\tt PEX} model. The mean $E_{\rm cut}$ in the entire \citet{Akylas2021} sample is therefore $E_{\rm cut,mean,low-z}$ = 204$\pm$31~keV (the uncertainties are at 1~$\sigma$). Such result is consistent with previous measurements with BeppoSAX \citep[covering 0.1--200~keV but has a lower sensitivity compared to NuSTAR][]{Boella1997} and NuSTAR \citep[e.g.,][]{Perola2002,Risaliti2002,Ricci2018,Tortosa2018,Balokovic2020,Kamraj2022}. }

{The mean $E_{\rm cut}$ in the PACHA sample is $E_{\rm cut,mean,high-z}$ = 101$\pm$21~keV. It is worth noting that the PACHA sample contains only 13 sources, so the use of Kaplan-Meier estimate should be interpreted with appropriate caution.}

We compute the mean values of cutoff energy in different bins of luminosity, black hole mass, and Eddington ratio, which are reported in Table~\ref{Table:mean}. We use the black hole mass and Eddington ratio measured in \citet{Koss2022}, which report M$_{\rm BH}$ and $\lambda_{\rm Edd}$ of 91 AGN in the \citet{Akylas2021} sample. Their cutoff energy as a function of the luminosity, black hole mass, and Eddington ratio and their means are plotted in Fig.~\ref{fig:lum_Edd_cut}. 

We further investigated the dependence of the cutoff energy on luminosity, black hole mass, and Eddington ratio using the generalized Kendall's $\tau$ survival-analysis test implemented in the ASURV package. The analysis indicates tentative anti-correlations between the cutoff energy and both luminosity ($p$ = {0.025$_{-0.019}^{+0.057}$}) and black hole mass ($p$ = {0.044$_{-0.034}^{+0.10}$}), whereas no statistically significant correlation is found with Eddington ratio ($p$ = {0.43$_{-0.25}^{+0.34}$}). 

It is worth noting that the high-$z$ AGN sample significantly expands the parameter space, e.g., the luminosity and black hole mass ranges. Without the high-$z$ AGN sample, the data from the local AGN do not reveal a detectable correlation between cutoff energy and luminosity ($p \sim 0.15$) or black hole mass ($p \sim 0.48$). 

\begin{figure*}[t] 
\centering
\includegraphics[width=.99\textwidth]{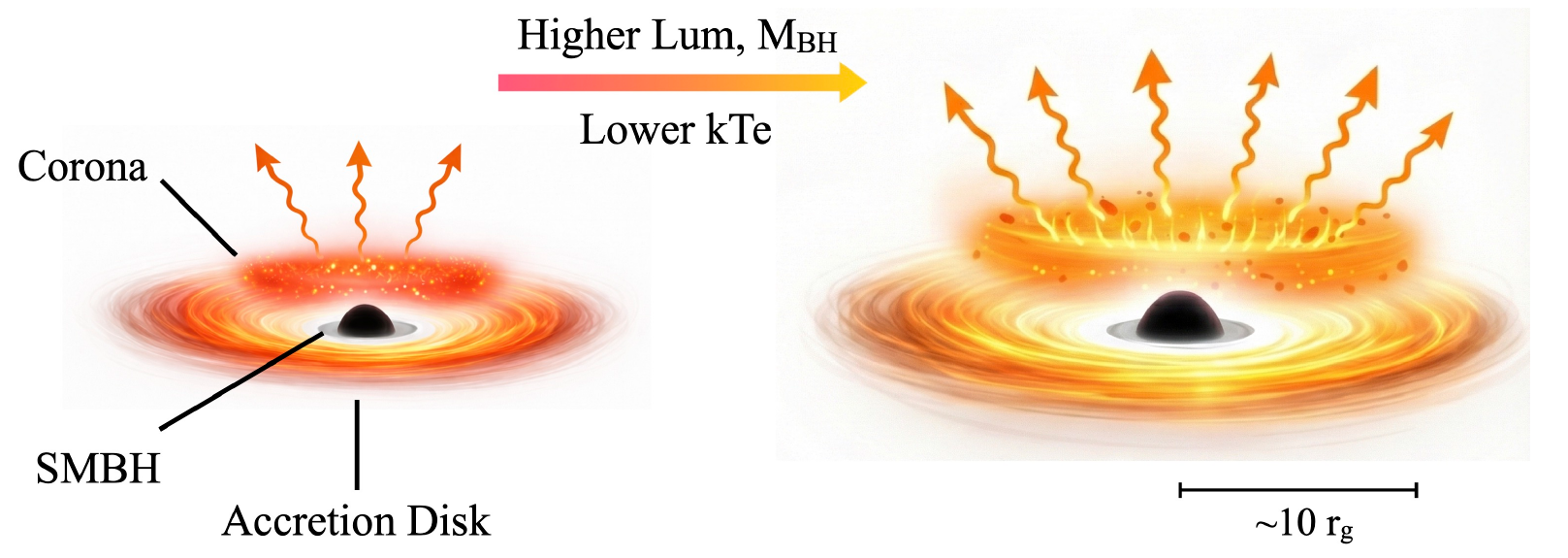}
\caption{Illustration of the coronal properties derived in this work. Sources with higher X-ray luminosities and larger black hole masses exhibit systematically lower coronal temperatures. The sizes of the various components are not to scale.}
\label{fig:illustration}
\end{figure*}  

We note that the X-ray luminosity and black hole mass in both the \citet{Akylas2021} sample and our PACHA sample are strongly correlated (Fig.~\ref{fig:BH_lum} in \ref{sec:BH_lum}), in the sense that more luminous sources tend to be associated with more massive black holes. Consequently, the observed dependence of the coronal high-energy cutoff could, in principle, be driven by either X-ray luminosity or black hole mass. Disentangling these two effects will require additional observations of luminous, low black hole mass AGN, as well as low luminosity, high black hole mass systems. 

Radiation MHD simulations of accretion disks indicate only a weak dependence of the coronal temperature on black hole mass \citep{Jiang2019}; see Section~\ref{sec:GRMHD} for further discussion. This suggests that X-ray luminosity is more likely to be the primary driver of the observed variations in coronal temperature.

We test for correlations between the photon index and reflection scaling factor and the intrinsic luminosity, black hole mass, and Eddington ratio in both the local and high-$z$ samples. The mean values of $\Gamma$ and $R$ in different $L_{\rm 2-10}$, $M_{\rm BH}$, and $\lambda_{\rm Edd}$ bins are reported in Table~\ref{Table:mean}. We do not observe significant evolution of either the photon index or the reflection scaling factor as a function of $L_{\rm 2-10}$, $M_{\rm BH}$, or $\lambda_{\rm Edd}$. The broadly similar spectral shapes (except for the high-energy cutoff) across AGN spanning a wide range of luminosities, black hole masses, and accretion rates suggest that the coronae may self-regulate to preserve a nearly uniform spectral shape.

A small number of luminous ($\log L_{\rm 2-10} > 45$), radio-quiet AGN at $z<1$ have previously been observed with \NuSTAR\ and exhibit well-constrained high-energy cutoffs \citep[e.g.,][]{Kammoun2016,Marinucci2022,Kammoun2023b}, consistent with the coronal properties measured in our PACHA sample. However, since these sources are at low redshift ($z<1$), they are not included in the present work. We will incorporate these objects in a forthcoming study that uniformly analyzes all type-1, radio-quiet AGN observed by \NuSTAR\ across a wide redshift range using the same coronal model, enabling tighter constraints on the potential correlations identified here.

{\citet{Peluso2026} studied a sample of 33 predominantly radio-quiet AGN spanning redshifts of $0.1<z<1$ and luminosities of $43.5<\log(L_{\rm 2-10})\lesssim45.5$, placing their sample between the PACHA and \citet{Akylas2021} samples in both redshift and luminosity. They measured a median cutoff energy of $\sim$90~keV for their intermediate-luminosity sample, significantly lower than that of the local, lower-luminosity AGN. By combining their sample with the local AGN sample, they found significant correlations between the cutoff energy and both the bolometric luminosity and Eddington ratio ($p<10^{-3}$), as well as a weaker correlation with black hole mass ($p\sim5\times10^{-3}$). These results are consistent with our findings, although they examined the dependence of the cutoff energy on bolometric luminosity, whereas we focus on the X-ray luminosity.}

\begingroup
\renewcommand*{\arraystretch}{1.2}
\begin{table*}
\begin{center}
\caption{The mean of cutoff energy, photon index, and reflection scaling factor in different 2--10~keV luminosity, black hole mass, and Eddington ratio bins. The error bars on the mean is reported at 1$\sigma$ confidence level. }\label{Table:mean}
  \begin{tabular}{cccccccc}
       \hline
       \hline
    &Bin&$N$&$E_{\rm cut,mean}$&$\Gamma_{\rm mean}$&$R_{\rm mean}$\\ 
    &&&keV\\
        \hline
        \hline
	log(L$_{\rm 2-10,low-z}$)&42 -- 43&27&{208$\pm$50}&{1.78$\pm$0.02}&{0.82$\pm$0.05}\\
	&43 -- 44&51&{208$\pm$38}&{1.81$\pm$0.01}&{0.78$\pm$0.05}\\
	&44 -- 45&32&{132$\pm$11}&{1.76$\pm$0.01}&{0.55$\pm$0.04}\\
        \hline
        	log(L$_{\rm 2-10,high-z}$)&45 -- 46&12&{94$\pm$23}&{1.86$\pm$0.01}&{1.01$\pm$0.15}\\
        \hline
        \hline
        	 log(M$_{\rm BH,low-z}$)&6 -- 7&16& {234$\pm$54}&{1.81$\pm$0.02}&{0.82$\pm$0.07}\\
	&7 -- 8&41&{215$\pm$50}&{1.80$\pm$0.01}&{0.72$\pm$0.04}\\
	&8 -- 9&28&{142$\pm$15}&{1.78$\pm$0.02}&{0.66$\pm$0.06}\\
        \hline
        	 log(M$_{\rm BH,high-z}$)&9 -- 10&13&{101$\pm$21}&{1.85$\pm$0.01}&{0.94$\pm$0.14}\\       
        \hline
        \hline
        	 log($\lambda_{\rm Edd,low-z}$)&-2.5 -- -1.5&15&{216$\pm$49}&{1.80$\pm$0.03}&{0.94$\pm$0.14}\\
	&-1.5 -- -1&39&{183$\pm$24}&{1.79$\pm$0.01}&{0.79$\pm$0.04}\\
	&-1 -- 0&33&{161$\pm$17}&{1.79$\pm$0.01}&{0.58$\pm$0.05}\\
        \hline
        	 log($\lambda_{\rm Edd,high-z}$)&-1 -- 0&11&{102$\pm$23}&{1.82$\pm$0.01}&{0.85$\pm$0.15}\\       
        \hline
         \hline     
\end{tabular}
\end{center}
\end{table*}
\endgroup

\subsection{Pair Regulation in AGN Coronae}
\lc{The heating and cooling of AGN coronae remain long-standing open questions. Direct Coulomb energy exchange between protons and electrons is generally too inefficient at the low densities inferred for AGN coronae to sustain the observed electron temperatures.}
\lc{The dissipation of magnetic energy, mediated by processes such as magnetic reconnection \citep{Rowan2017} and turbulence \citep{Comisso18}, is widely regarded as the primary mechanism for electron energization, although the detailed microphysics is still uncertain.}
\lc{On the other hand, the compact size of the corona combined with its high luminosity implies highly efficient radiative cooling, dominated by inverse-Compton scattering of soft seed photons by energetic electrons.}

The key parameters to probe the heating and cooling process of the compact corona are the coronal temperature and the compactness. The compactness is defined as 
\begin{equation}
\ell = \frac{L_c}{R_c}\frac{\sigma_T}{m_ec^3}
\end{equation}

\noindent where $L_c$ is the coronal luminosity from 0.1 to 200~keV (which captures the majority of the coronal emission), $R_c$ is the coronal radius assuming a spherical geometry, $\sigma_T$ is the Thomson cross-section, and $m_e$ is the mass of the electron (511~keV/$c^2$). The equation can also be expressed as

\begin{equation}
\ell = 4\pi\frac{L_c}{L_{\rm Edd}} \frac{m_p}{m_e}\frac{R_g}{R_c}
\end{equation}
where $L_{\rm Edd}$ is the Eddington luminosity ($L_{\rm Edd}$ = 1.26$\times$10$^{38}$ M$_{\rm BH}$/M$_\odot$ erg~s$^{-1}$) and $m_p$ is the mass of the proton (938~MeV/$c^2$). We assume coronae of all sources to be spherical with a radius of 10~$r_g$, which is approximately the median of the coronal size of the AGN measured in a sample of gravitationally lensed AGN \citep[\ref{sec:mirco} and][]{Chartas2016}. The estimated compactness of each source is reported in Table~\ref{Tab:source}.

At high compactness ($\ell\gg1$), the Compton cooling time of an electron ($t_{\rm C}$) is much \lc{shorter than the coronal light-crossing time \citep[$t_{\rm cross}$,][]{Fabian2015}, with 
\begin{equation}
\frac{t_{C}}{t_{cross}} = \frac{3\pi}{2\ell(1+\tau)} \, .
\end{equation}
This indicates that radiative cooling is extremely efficient, with electrons losing energy on timescales well below the coronal light-crossing time.}

\lc{When a sufficient fraction of photons attain high energies, either due to elevated coronal temperatures or high compactness, photon--photon interactions can produce electron–positron pairs once their center-of-mass energy exceeds $\sim 1$~MeV.}
\lc{The production of cool pairs increases the lepton number density in the corona and at fixed heating power, this leads to a reduction in the mean energy per lepton.}
Therefore, the pair-production process can act as a thermostat: increases in coronal temperature or compactness trigger enhanced pair production, which in turn regulates and limits the coronal temperature. The resulting limiting temperature for a given compactness, which depends on the coronal geometry, the properties of the input soft photons, and the distribution of electron energies, is commonly referred to as the pair-line.

\begin{figure}[t] 
\centering
\includegraphics[width=.49\textwidth]{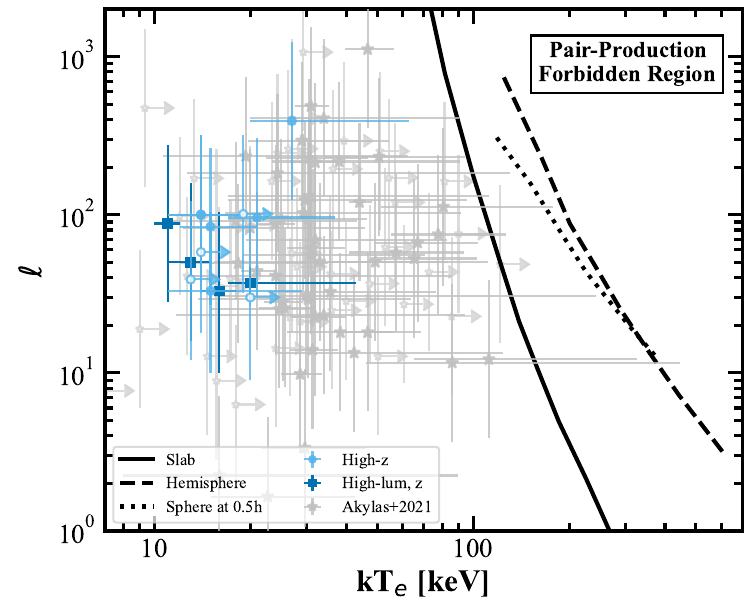}
\caption{High-$z$, high-luminosity AGN studied in this work (blue) and selected {\it Swift}-BAT AGN (grey) from \citet{Akylas2021} on the $kT_{\rm e}$--$\ell$ plane. The cutoff energy measured in \citet{Akylas2021} is converted to the coronal temperature assuming $E_{\rm cut}$ = 3~$kT_{\rm e}$. The pair-lines assuming geometries of slab (solid), hemisphere (dashed), and a sphere at a height equal to half the radius of the sphere (dotted) adopted from \citet{Stern1995} and \citet{Svensson1996}.}
\label{fig:kTe_l}
\end{figure}

In Fig.~\ref{fig:kTe_l}, we compare our sources with the pair lines of different coronal geometries of slab, hemisphere, and a sphere at a height equal to half the radius of the sphere \citep{Stern1995,Svensson1996}. The mean coronal temperature of our entire PACHA sample using the Kaplan-Meier estimator is $kT_{\rm e}$ = {21.4$\pm$2.9}. Besides the four sources with lower limits on their coronal temperature due to their low-flux state, the remaining nine sources all have much lower coronal temperature compared to the pair-lines for different geometries, even considering the large uncertainties on compactness. Therefore, the result is in good agreement with the pair-production model, which predicts that stable coronal configurations cannot exist within the pair-production runaway forbidden region, i.e., above (to the right of) the pair-lines in the $kT_{\rm e}$--$\ell$ plane, where runaway pair production would occur.

Agreement with the pair-line is also found for the local {\it Swift}-BAT sample. We plot sources with constrained cutoff energy in the {\it Swift}-BAT sample in Fig.~\ref{fig:kTe_l}. The cutoff energy measured in \citet{Akylas2021} is converted to coronal temperature using the relation $E_{\rm cut}$ = 3~$kT_{\rm e}$ \citep{Petrucci2001} for optically-thick corona, as all of the sources in the {\it Swift}-BAT sample have coronal optical depths of $\tau > 1$ (Fig.~\ref{fig:tau}). {Nevertheless, we caution that the conversion factor between the cutoff energy and coronal temperature can vary substantially, with $E_{\rm cut}/kT_{\rm e}$ ranging from $\sim$2 to 5, particularly for high-temperature coronae \citep{Middei2019}. We therefore plan to reanalyze the local AGN sample using the same Comptonization model in future work (Granadoz et al., in prep.) to ensure a consistent comparison.} 

The estimated mean coronal temperature of the sources with constrained cutoff energy is $kT_{\rm e}$ = 68$\pm$10~keV. We caution that the conversion factor between coronal temperature and high-energy cutoff can still vary for individual sources. A detailed spectral analysis of the sources in \citet{Akylas2021} using physical Comptonization models would therefore be required to obtain a more reliable estimate of the mean coronal temperature for their sample.
The compactness parameters are calculated using the 0.1--200 keV luminosities, which are converted from the observed 2--10 keV band reported in \citet{Akylas2021} assuming an universal conversion factor of $cf$ = {4.5}. The conversion factor is derived by adopting a spectral model with parameters fixed to the mean values of the entire sample, i.e., E$_{\rm cut,mean}$ = {200~keV} and $\Gamma_{\rm mean}$ = 1.79. A coronal radius of 10~$r_g$ is assumed. 

We find that no AGN (including those with only lower limits on their high-energy cutoff) has been robustly confirmed to reside within the pair-production runaway region, in agreement with the pair-regulation model. 

It is worth noting both AGN in our PACHA sample and in the {\it Swift}-BAT selected sample exhibit comparable compactness values, spanning $\ell\sim$10--500 (assuming a coronal radius of 10~$r_g$). In this regime, the coronae are dominated by pairs \citep[Fig.~5 in][]{Fabian2017}, where the observed coronae are expected to cluster near the pair-lines. However, all AGN with $\ell>10$ and well-constrained cutoff energies, comprising $\sim$50\% of the sample in this work, lie significantly below the theoretical pair lines. These results cannot be reconciled with the above pair-production model that assumes a single-temperature, purely thermalized corona.

\begin{figure*}[t] 
\centering
\begin{minipage}{.49\textwidth}
\includegraphics[width=.99\textwidth]{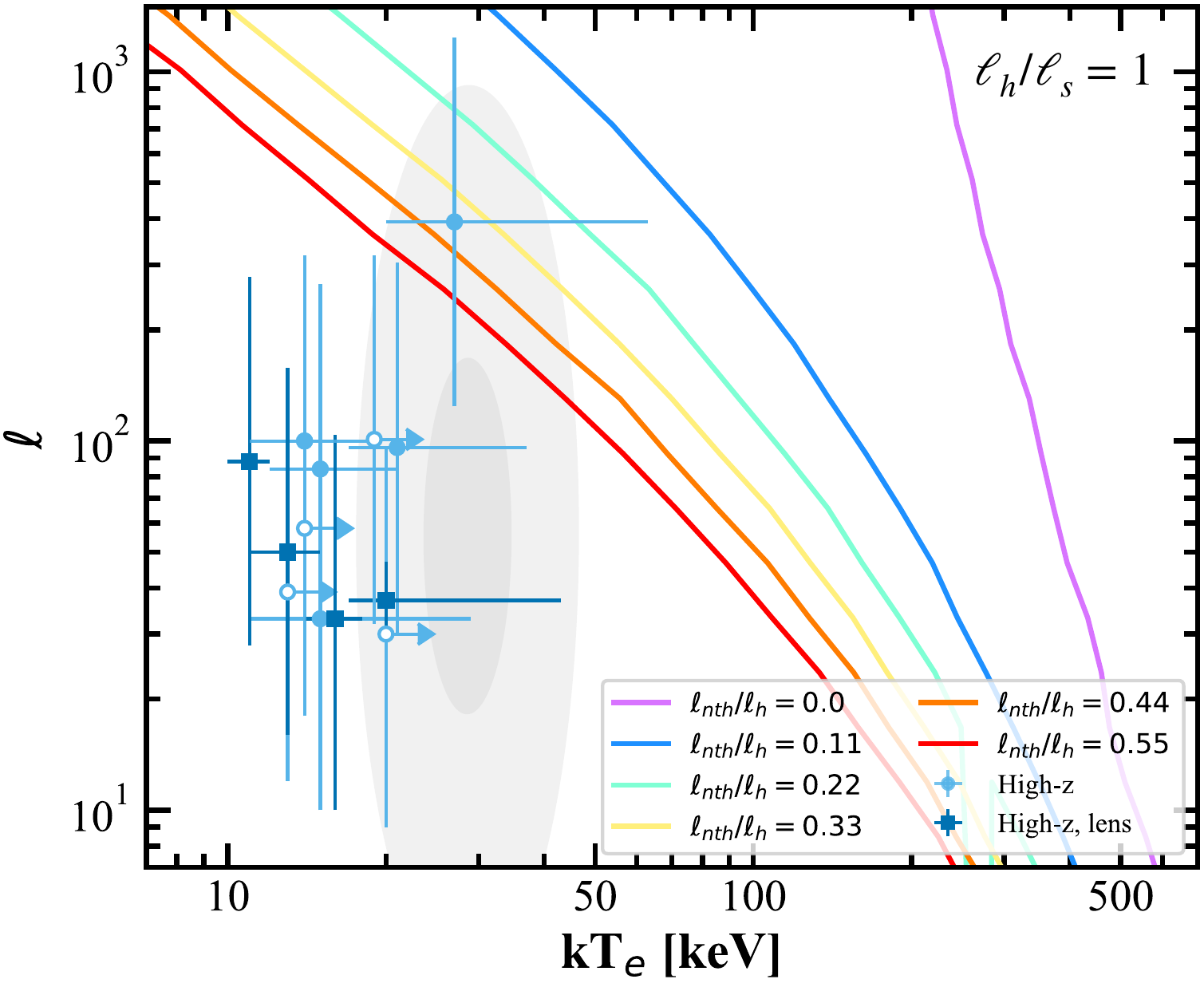}
\end{minipage}
\begin{minipage}{.49\textwidth}
\includegraphics[width=.99\textwidth]{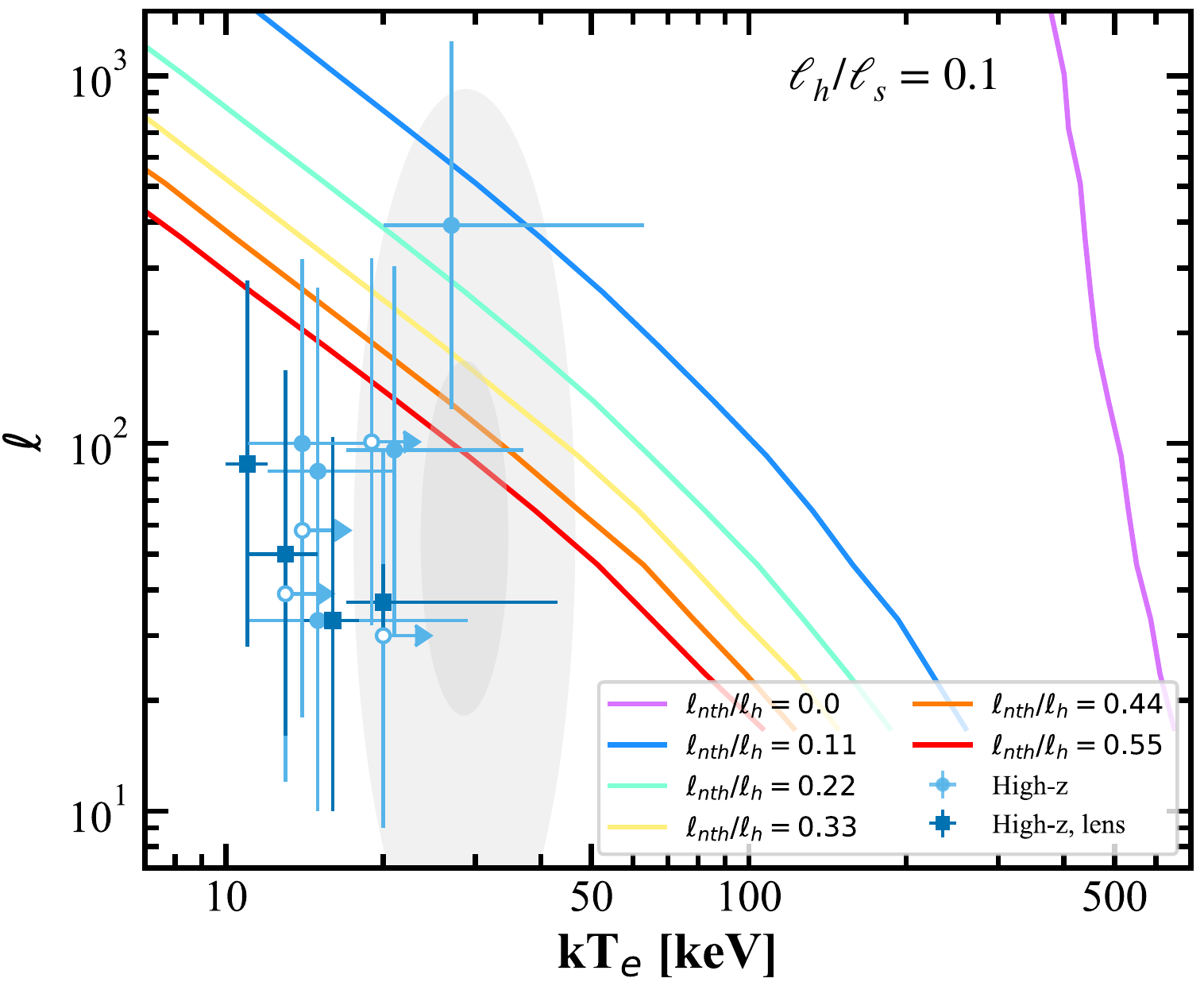}
\end{minipage}
\caption{Sources studied in this work on the $kT_{\rm e}$--$\ell$ plane under a hybrid-corona model with a spherical corona geometry assuming $\ell_{\rm h}$/$\ell_{\rm s}$ = 1 (left) and $\ell_{\rm h}$/$\ell_s$ = 0.1 (right). The pair-equilibrium lines from right to left are $\ell_{nth}$/$\ell_{\rm h}$ = 0.0 (purple), $\ell_{\rm nth}$/$\ell_{\rm h}$ = 0.11 (blue), $\ell_{\rm nth}$/$\ell_{\rm h}$ = 0.22 (cyan), $\ell_{\rm nth}$/$\ell_{\rm h}$ = 0.33 (yellow), $\ell_{\rm nth}$/$\ell_{\rm h}$ = 0.44 (orange), and $\ell_{nth}$/$\ell_{\rm h}$ = 0.55 (red). The grey contours comprise 30\% and 75\% of the selected {\it Swift}-BAT AGN with cutoff energy constrained in \citet{Akylas2021}. A coronal size of 10~$r_{\rm g}$ is assumed for both samples.}
\label{fig:hybrid}
\end{figure*}   
\subsection{Electron Distributions in AGN Coronae}

The low-coronal-temperature is proposed to be explained by a hybrid-corona model \citep{Ghisellini1993,Zdziarski1993,Zdziarski1996}. In this model, coronae include a major population of thermal electrons under a Maxwellian distribution combined with a small hard tail of non-thermal electrons \citep[see, e.g., Fig.~2 in][]{Fabian2017}. The small fraction of high-energy, non-thermal electrons produce a large population of pairs, which effectively cools the temperature of the thermal electron population in the corona. This model predicts that we would measure a low coronal temperature when fitting X-ray spectra below $\sim$100~keV, primarily produced by thermal electrons, while the hard-tail of non-thermal electrons would require sensitivity at a few hundreds of keV \citep[see, e.g., Fig.~4 in][]{Fabian2017} and are not expected to be detectable by current telescopes \citep{Gondek1996}, if they exist. \citet{Fabian2017} found that many of the low coronal temperatures observed in a dozen local AGN can be explained by a small fraction (10--30\%) of non-thermal electrons. \lc{Such a hybrid electron distribution can arise naturally from magnetic energy dissipation via turbulence and reconnection, which efficiently energize $\sim$ 10-30\% of the electrons into a non-thermal tail \citep{Comisso19}.}

Including the non-thermal electrons will significantly enhance pair-production. Therefore, this hybrid corona model predicts a set of pair-equilibrium lines (with different non-thermal electron fractions) far below the pair-lines predicted in Fig.~\ref{fig:kTe_l}. We compare our high-luminosity AGN sample and the local, low-luminosity AGN sample with hybrid-corona models assuming different non-thermal electron fractions, \((\ell_{\rm nth}/\ell_{\rm h})\), following \citet{Fabian2017}, where \(\ell_{\rm nth}\) and \(\ell_{\rm h}\) denote the compactness of non-thermal electrons and the total coronal heating power, respectively. For approximately half of the local, low-luminosity AGN in the \citet{Akylas2021} sample whose coronal temperatures are unconstrained, the observed limits are consistent with models invoking a non-thermal electron fraction of \(\lesssim 55\%\). 

However, for the majority of sources with constrained coronal temperatures, including both the local, low-luminosity AGN from \citet{Akylas2021} and the high-redshift, high-luminosity AGN in our sample, the measured temperatures at a given compactness are significantly lower than predicted by the hybrid-corona model, even when assuming a non-thermal fraction of \(\sim 55\%\) (Fig.~\ref{fig:hybrid}, left panel). This discrepancy indicates that variations in the non-thermal electron fraction alone are insufficient to account for the low coronal temperatures observed in these AGN.
  
In the hybrid-corona scenario, the location of the pair-equilibrium lines in the $kT$--$\ell$ plane also depends on the ratio between the total coronal heating power and the soft-photon compactness, $\ell_h/\ell_s$. A value of $\ell_h/\ell_s = 1$ approximately corresponds to comparable heating and seed-photon power intercepted from the accretion disk, whereas $\ell_h/\ell_s \ll 1$ implies more efficient radiative cooling owing to an enhanced soft-photon field. The pair-equilibrium lines shown in the left panel of Fig.~\ref{fig:hybrid} adopt $\ell_h/\ell_s = 1$. Reducing the ratio to $\ell_h/\ell_s = 0.1$ shifts the predicted equilibrium to lower coronal temperatures at fixed compactness (right panel of Fig.~\ref{fig:hybrid}; \citealt{Fabian2017}). Nevertheless, even allowing for substantial non-thermal heating fractions (up to $\ell_{\rm nth}/\ell_h \approx 0.55$), the models still struggle to reproduce the lowest coronal temperatures observed in a subset of AGN with well-constrained measurements.

Therefore, the low coronal temperature of these AGN suggests a combination of a high non-thermal electron fraction ($\ell_{\rm nth}/\ell_{\rm h}>33\%$) and highly efficient Compton cooling ($\ell_{\rm h}/\ell_{\rm s}<0.1$) of the corona. Furthermore, the coronal temperature of high-luminosity AGN is systematically lower than those of the low-luminosity AGN, suggesting a requirement of either higher non-thermal electron fraction and/or a lower heating to soft photon power ratio for high-luminosity AGN. Indeed, the bolometric to X-ray luminosity ratio (which is a proxy of the heating to soft photon power ratio) is much higher for high-luminosity AGN (see Section~\ref{sec:k-correction} for more details).

{However, it is also worth noting that a low $\ell_h/\ell_s$ may produce an extremely soft X-ray spectrum. For example, \citet{Coppi2000} found that $\ell_h/\ell_s \sim 0.1$ yields a photon index of $\Gamma \gtrsim 3$ (with very weak dependence on the non-thermal fraction) in their hybrid coronal model. Since such steep spectra are generally not observed in our sources, additional physical processes regulating the conversion of soft-photon energy into hard X-ray emission may be required.}

The pair lines are computed assuming a spherical coronal geometry in \citet{Fabian2017}. Recently, the Imaging X-ray Polarimetry Explorer \citep[IXPE;][]{ixpe22} has observed a subset of bright, local, type-1 AGN with the aim to constrain their coronal geometry \citep[e.g.,][]{Marinucci2022b,Gianolli2023, Gianolli2024,Tagliacozzo2023, Ingram2023, Pal2025}. These works suggest an asymmetric coronal geometry. However, in most cases the observations resulted in an upper limit on the X-ray polarization fraction, leading to poor constraints on the coronal geometry. In the case of NGC 4151, \cite{Gianolli2023, Gianolli2024} reported the detection of a polarization degree of $4.5 \pm 0.9\%$ with a polarization angle parallel to the extended radio emission in the source. This indicates that the source of polarized X-rays is located in the plane normal to the jet, which could hint at a radially extended corona. Alternative scenarios in which the polarized X-rays are produced by accretion disk reflection with a compact, centrally located corona can also explain the IXPE results (Kammoun et al., in prep.) 

Alternative geometries (e.g., slab-like geometry; see Fig.~\ref{fig:kTe_l}) may account for the low coronal temperatures observed in both high-$z$ and local AGN without requiring such extreme regions of parameter space, a scenario that is worth further investigation in future work. In addition, uncertainties in the coronal size (other than 10~$r_g$; see \ref{sec:mirco}) further complicate the determination of both the thermal electron fraction and the cooling efficiency.

The mean coronal optical depth of our entire PACHA sample using the Kaplan-Meier estimator is $\tau_{\rm high-z}$ = 4.8$\pm$0.2. The optical depth of the \citet{Akylas2021} sample is $\tau_{\rm low-z}$ = {2.4$\pm$0.1}. Within a hybrid coronal framework, {enhanced electron--positron pair production rate increases the total lepton density in the corona, thereby increasing its optical depth.} An increasing optical depth will lead to a lower coronal temperature at fixed $\ell_{\rm h}$, as the radiative cooling (through scattering) becomes more efficient while the heating power remains unchanged. However, it is not clear whether variations in $\tau$ alone are enough to account for the full range of observed coronal temperatures.

\subsection{Interpretation from Radiation MHD Simulations} \label{sec:GRMHD}
\lc{The electron temperatures inferred from our measurements ($\sim$10–40 keV), together with the observed high-energy cutoffs ($E_{\rm cut}\sim$50–150 keV), are broadly consistent with the thermal properties of coronal gas produced in state-of-the-art global radiation MHD simulations of accretion flows \citep{Jiang2019}.
In these simulations, electrons and ions are assumed to be thermally coupled and share a single temperature, which is determined self-consistently by the balance between turbulent heating driven by magnetorotational instability and radiative cooling computed via radiation transport.}

For Eddington ratios of $7\%-20\%$ and a $5\times 10^8$ solar mass black hole, \lc{these simulations predict} gas temperatures at the optically thick disk mid-plane \lc{of order} $10^5$ K, \lc{rising rapidly to} $10^8-10^9$ K ($\sim10-100$ keV) in the region of moderate optical depth $\tau \sim 1-5$ (i.e., the coronal region) where radiative cooling is inefficient. Magnetic buoyancy carries the magnetic energy towards the disk surface and brings a significant fraction of the accretion power to be dissipated in the coronal region, producing X-rays. The magnetic dissipation fraction decreases as accretion rate increases, leading to systematically lower coronal temperatures.

The disk thickness also increases with increasing accretion rates, leading to higher densities in the corona region. This enhances radiative cooling and consequently lowers the gas temperature. These trends are qualitatively consistent with the observational results presented here. Since these simulations assume a purely thermal electron population, this may indicate that non-thermal electrons might not be the primary drivers of the observed thermal properties of coronae.

For a fixed Eddington ratio, the effective temperature of the accretion disk scales as $M_{\rm BH}$$^{-0.25}$ in the standard $\alpha$-disk model \citep{Shakura1973}. However, the measured coronal temperature shows a much weaker dependence on black hole mass (Table~\ref{Table:mean}). This weak scaling suggests that the thermal properties of the corona are not strongly coupled to the accretion disk.

{Our results are also consistent with other general relativistic magnetohydrodynamic (GRMHD) simulations, which predict similar coronal temperatures over a comparable range of black hole masses and accretion rates \citep[e.g.,][]{Nagele2026}.}

\subsection{General Relativistic Effects} \label{sec:GR}
AGN coronae are believed to be compact and reside in the immediate vicinity of the central SMBHs \citep[see][and references therein]{Reis2013}. Consequently, coronae reside deep within the strong-gravity regime of SMBHs where relativistic effects such as gravitational redshift, Doppler boosting, and light bending can significantly affect observed coronal properties.

\begin{itemize}
\item {Gravitational Redshift.} The intrinsic X-ray spectrum of the corona can be redshifted due to the effects of strong gravitational, as the corona is very close to the SMBH. The observed cutoff energy of the spectrum is $E_{\rm cut,obs}$ = $E_{\rm cut,int}$/(1 + $z_g$), where $z_g$ is the gravitational redshift at the characteristic emission radius ($r$) to the SMBH. The intrinsic luminosity is also affected through Liouville’s theorem with $L_{\rm obs}$ = $L_{\rm int}$/(1 + $z_g$)$^4$. For a Schwarzschild black hole, the gravitational redshift is 1 + $z_g$ = (1 -- 2$r_g$/$r$)$^{-1/2}$. 
The characteristic emission radius is largely uncertain and may depend on the coronal geometry. Assuming $r$ = 10~$r_g$, the gravitational redshift leads to a $E_{\rm cut,int}\sim1.1\,E_{\rm cut,obs}$ and $L_{\rm obs}$ = 0.64\,$L_{\rm int}$. The inferred coronal temperature is likewise affected. For a rotating SMBH, the gravitational redshift depends not only on radius but also on black hole spin and inclination, which can affect the observed cutoff energy. 

\item {Doppler Boosting.} When the electron plasma in coronae possesses bulk motion toward the observer \citep[e.g.,][]{Ewing2025}, relativistic Doppler effects amplify the observed luminosity and frequency of the X-ray photon. The enhancement factor for luminosity is given by $\mathcal{D}^{3+\alpha}$, where $\alpha$ is the spectral index of the emission and $\mathcal{D}$ is the Doppler factor: $\mathcal{D}$ = 1/$\gamma$/(1--$\beta$cos$\theta$) with $\beta = v/c$, $\gamma = 1/\sqrt{1 - \beta^2}$, and $\theta$ the angle between the velocity vector and the observer’s line-of-sight. The frequency transformation is $h\nu_{\rm obs} = \mathcal{D}\,h\nu_{\rm int}$. Therefore, the observed cutoff energy is $E_{\rm cut,obs}$ = $\mathcal{D}\,E_{\rm cut,int}$. For example, a corona moving outward at $v = 0.1\,c$ along the line-of-sight with $\alpha = 1$ yields a luminosity enhancement of $\sim$49\% and $E_{\rm cut,obs}\sim1.1\,E_{\rm cut,int}$ for a face-on inclination. The disk inclination is likely to be almost face-on in these type 1 quasars, suggesting a small rotational Doppler shift. 

\vspace{0.1cm}

\item {Light Bending.} In curved spacetime, photon trajectories are bent toward the black hole and the accretion disk. As a result, light bending enhances the fraction of intrinsic coronal emission reflected by the disk, with the magnitude of the effect depending on the coronal distance from the SMBH and the viewing inclination. Consequently, the observed coronal luminosity can be reduced relative to the intrinsic coronal luminosity.

\end{itemize}

In summary, relativistic effects introduce complex distortions to both the intrinsic cutoff energy (and hence the coronal temperature) and the luminosity in the immediate vicinity of the SMBH.

\begin{figure*}[t] 
\centering
\includegraphics[width=.9\textwidth]{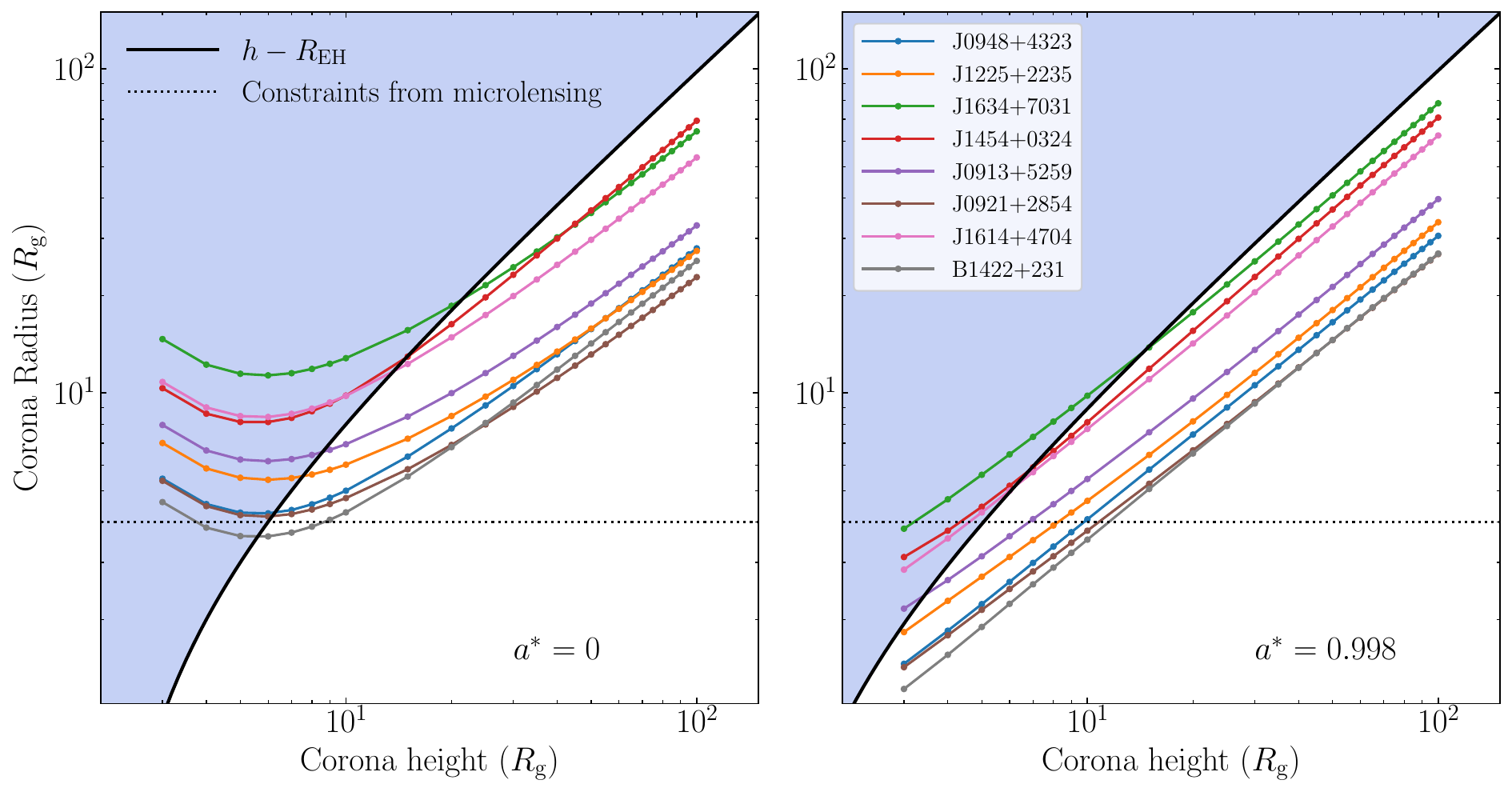}
\caption{Coronal height ($h_c$) as a function for coronal radius ($R_c$) of the sources in our sample estimated using the method in \citet{Dovciak22} assuming black hole spin $a^\ast$ = 0 (left) and $a^\ast$ = 0.998 (right). Coronal height should be larger than the coronal radius, thus providing a forbidden (blue) region. The solid line $h$--$R_{\rm EH}$ is the coronal height subtracted by the event horizon radius of the SMBH. The lowest coronal size measured from micro-lensing is indicated with a horizontal dashed line \citep{Chartas2016}.}
\label{fig:R_h}
\end{figure*} 
\citet{Tamborra2018} quantified GR effects on the measurement of the cutoff energy near an SMBH, including gravitational and Doppler energy shifts as well as light bending. They defined a cutoff-energy correction factor as the ratio between the intrinsic and observed cutoff energies, $g = E_{\rm cut}^{\rm i}/E_{\rm cut}^{\rm o}$. For a lamp-post geometry, the $g$-factor primarily depends on the coronal height $h$. Assuming a characteristic coronal size and height of $10\,r_g$, they found $g \sim 1.1$ for both Kerr and Schwarzschild black holes. For an extended disk corona, $g$-factor depends on both the inclination angle and the outer coronal radius $r_{\rm out}$. For $r_{\rm out} = 10\,r_g$ and a face-on configuration, $g \approx 1.3$ for both Kerr and Schwarzschild metrics, while a more extended corona ($r = 20\,r_g$) yields a smaller correction, $g \approx 1.2$. 

These results indicate that GR corrections to the measured cutoff energy are modest, particularly for extended coronae. Therefore, GR effects are unlikely to alter our conclusion that high luminosity, high black hole mass AGN exhibit lower coronal temperatures. Moreover, since GR corrections apply to both the local comparison sample and the PACHA sample, and may even be more pronounced in local, lower $M_{\rm BH}$ sources as their coronae might be more compact (Table~\ref{Table:CS_micro}). Given that the impact on the high-energy cutoff and coronal temperature is modest, we did not apply relativistic corrections to either our measurements or those reported in \citet{Akylas2021}.

\subsection{Constraints on Coronal Height and Size}
Determining the corona’s location, size, and geometry is crucial for understanding the physical conditions and energy release in the innermost accretion flow. It also provides a direct, quantitative link between observed X-ray reflection/reverberation signatures \citep[e.g.,][]{Demarco2013, Kara16} and the underlying accretion physics. Sizes estimated from micro-lensing of high-redshift quasars suggest that X-ray coronal sizes are of the order of a few to tens of gravitational radii \citep[e.g.,][]{Chartas2016}. Coronal heights are better estimated in the local universe using reverberation studies and also suggest heights of the order of tens of gravitational radii \citep[see][]{Demarco2013, Kara16, Kammoun2023, Papoutsis2024}. It is worth noting that the corona is expected to be dynamic and several works have shown hints of changes in the coronal size and/or location \citep[see e.g.,][]{Wilkins2015, Wilkins2022, Kammoun2024, Papoutsis2025}.

The data quality of our PACHA sample does not allow us to obtain precise estimates of the location and size of the X-ray coronae. However, we are able to estimate a physical range of coronal sizes and heights under the assumption of a spherical corona located on the rotational axis of the black hole. To do so, we use the \texttt{KYNSED} code \citep{Dovciak22}. \texttt{KYNSED} estimates the coronal radius by modeling the corona as a Comptonizing region that up-scatters thermal seed photons from the accretion disc, and then uses energy and photon-number conservation to infer the corona's effective cross-sectional area (and hence its radius). First, \texttt{KYNSED} computes the disk thermal luminosity and photon flux incident on the corona in the coronal rest frame, which sets the characteristic seed-photon energy entering the corona. Next, it imposes energy conservation: the total luminosity emerging from the corona equals the coronal heating power plus the energy carried by the fraction of seed photons that enter and are scattered, where this fraction is controlled by the coronal optical depth. In parallel, \texttt{KYNSED} imposes photon-number conservation for Comptonization. The number of photons in the emerging Comptonized spectrum equals the number of seed photons that enter the corona and scatter, which scales with the corona's cross-sectional area. Solving these constraints together yields the coronal area and thus the coronal radius under the assumption of a spherical corona.

For each of the sources (except the lensed ones), we use the black hole mass, photon index, X-ray luminosity, high-energy cutoff, and redshift listed in Table~\ref{Tab:source}, and we assume two values for the black hole spin $a^\ast =0$ (i.e., non-spinning) and $a^\ast =0.998$ (i.e., maximally spinning, at the Thorne limit). Then, we estimate the coronal radius from \texttt{KYNSED} by changing the height between 3\,$r_{\rm g}$ and 100\,$r_{\rm g}$. For each height we adjust the luminosity in the coronal rest-frame in a way that produces the same observed 2-10\,keV flux. Fig.~\ref{fig:R_h} shows the derived coronal radius as a function of height for the non-spinning and maximally spinning black hole scenarios (left and right panels, respectively). In this exercise, a solution is considered physical if, at a given height and spin, the coronal radius does not extend within the event horizon \citep[see][]{Dovciak22}. The shaded blue regions mark the non-physical solutions where the coronal radius is larger than the coronal height. We note that, for some sources, the scenario with a maximally spinning black hole allows very compact coronae (as small as $\sim 1-2$~$r_{\rm g}$) at very low heights (as low as 3~$r_{\rm g}$). For the non-spinning black hole scenario, the lowest allowed radius is $\sim 4$~$r_{\rm g}$ for $h\geq5$~$r_{\rm g}$. For both spins, the maximum allowed coronal radius is $\sim 70-80$~$r_{\rm g}$. The minimum/maximum inferred radii are not the same for all sources, thus showing a range of possible corona sizes. We stress that this exercise is not intended to provide precise estimates of the coronal radius or height. Our goal is to demonstrate that, using physically motivated models, the observed coronae agree with the assumption of compact sizes of tens of gravitational radii. The values we estimate here are in agreement with the values inferred from micro-lensing and reverberation studies.

\begin{figure*}[t] 
\centering
\begin{minipage}{.49\textwidth}
\includegraphics[width=.99\textwidth]{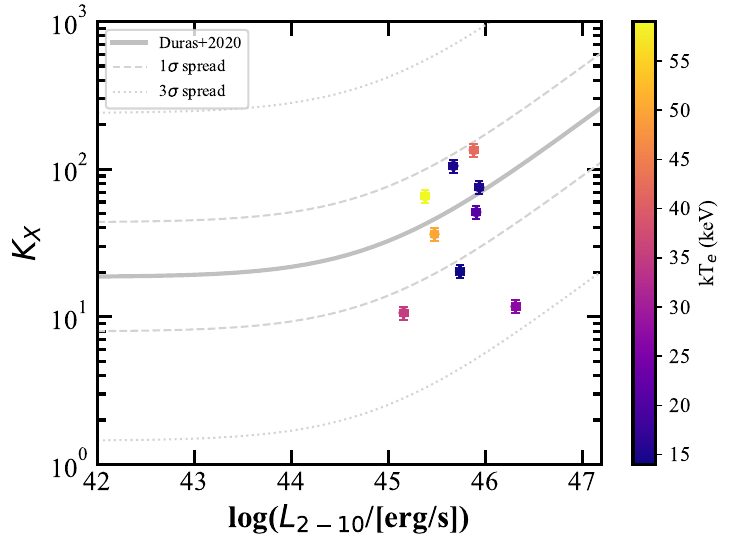}
\end{minipage}
\centering
\begin{minipage}{.49\textwidth}
\includegraphics[width=.99\textwidth]{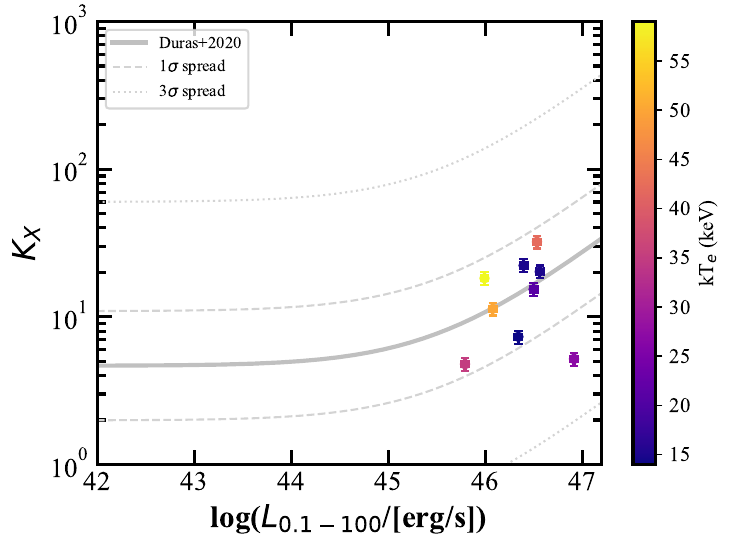}
\end{minipage}
\caption{Bolometric to X-ray luminosity ratio as a function of 2--10~keV (left) and 0.1--100~keV (right) luminosity of the AGN in the PACHA sample. Lensed sources are excluded because their bolometric luminosities cannot be derived in a self-consistent manner using the SED-fitting method applied to the non-lensed targets. The symbol color indicates the coronal temperature. The solid grey line is the $L_x$--$k_x$ relation derived in \citet{Duras2020} and the dashed and dotted lines are the 1$\sigma$ and 3$\sigma$ spreads.}
\label{fig:Lx_k}
\end{figure*}  
\subsection{Bolometric to X-ray Luminosity}\label{sec:k-correction}

The bolometric to X-ray luminosity ratio ($k_{\rm X}$ = $L_{\rm bol}$/$L_{\rm 2-10}$, where $L_{\rm bol}$ is the bolometric luminosity and $L_{\rm 2-10}$ is the 2--10~keV luminosity) is a significant diagnostic for the energy transfer process among different structures in AGN, including the corona and disk. The ratio reflects how efficiently the corona extracts and radiates the energy from accretion. Studying the correlation between $k_X$ and accretion rate (or Eddington ratio) can provide useful information on corona--disk coupling. Previous works on the bolometric to X-ray luminosity ratio have revealed that $k_X$ presents strong dependence on the AGN luminosity (both X-ray and bolometric) and Eddington ratio; specifically $k_x$ increases for luminous and highly-accreting AGN \citep{Elvis1994,Shankar2004,Vasudevan2007,Lusso2012,Netzer2019,Duras2020,Lpez2024}. This implies that the corona extracts and radiates less efficiently compared to the disk in luminous and highly-accreting state. Furthermore, the bolometric to optical luminosity ratio ($k_o$) is found to be nearly independent of luminosity and Eddington ratio. This intrinsic weak X-ray emission in powerful AGN might be caused by the stronger photon-trapping and advection of X-ray photons into the SMBH \citep[e.g.,][]{Leighly2007} or disk radiation pressure suppressing the coronal heating \citep{Arcodia2019} or powerful disk-winds \citep{Nardini2019,Zappacosta2020}. Global 3D radiation MHD simulations show that the disk becomes denser and more magnetically supported as accretion rate increases, so less power is dissipated in the optically thin (corona) region and the corona becomes more compact and intrinsically weaker \citep[e.g.,][]{Jiang2019}.

We derive the bolometric luminosity of the targets in our sample by fitting their spectral energy distribution (SED) from X-rays to radio utilizing the Code Investigating GALaxy Emission ({\tt CIGALE}) v2022.1 \citep{Boquien2019,Yang2022}. The {\tt CIGALE} package has been widely proved to well reproduce the SED of both galaxies and AGN \citep[e.g.,][]{Wang2019,Salim2020,Shirley2021,Lpez2024}. The detailed {\tt CIGALE} analysis is described in Section~\ref{sec:SED}. 

The bolometric luminosities of the AGN in our sample are integrated from 1~$\mu$m to $\sim$0.12~\AA\ (100~keV), and are reported in Table~\ref{Tab:source}. We did not perform SED fitting of the gravitationally lensed targets, as their SEDs can be distorted due to the differential magnification at different wavelengths. \citet{Duras2020} defines bolometric luminosities differently, by integrating from 1~$\mu$m to 20~\AA\ ($\sim$0.62~keV). Therefore, we add the 0.62--100~keV luminosity to their bolometric luminosity to take into account the X-ray contribution to bolometric luminosity of AGN. The 0.62--100~keV luminosity is computed by converting their 2--10~keV luminosity using a conversion factor of $cf_{\rm x}$ = 3.3, assuming a spectral shape of E$_{\rm cut,mean}$ = 200~keV and $\Gamma_{\rm mean}$ = 1.79.

We plot the bolometric to X-ray luminosity ratio as a function of 2--10~keV luminosity of the sources in our sample in Fig.~\ref{fig:Lx_k}. We compare it to the $L_X$--$k_X$ relation derived in \citet{Duras2020}, which comprised about 1,000 AGN across seven orders of magnitude of luminosity. The majority of the sources in our high-redshift sample indeed present intrinsically weaker X-ray emission compared to their bolometric emission, e.g., compared to the low-luminosity AGN. Nevertheless, three targets lie below the 1$\sigma$ spread of the \citet{Duras2020} $L_X$--$k_X$ relation, which is expected as we selected X-ray bright sources for \NuSTAR\ observations. 

Considering that the entire X-ray emission (e.g., 0.1--100~keV) would contribute more significantly than the 2--10~keV luminosity to the bolometric (1~$\mu$m to 100~keV) luminosity of AGN, we plot the bolometric to X-ray luminosity ratio ($k_{X,\rm full}$ = $L_{\rm bol}$/$L_{\rm 0.1-100}$) in Fig.~\ref{fig:Lx_k}. We convert the 2--10~keV to 0.1--100~keV luminosity of the \citet{Duras2020} relations using a conversion factor of $cf_{\rm x}$ = 4, assuming the above spectral shape. Therefore, the X-ray (0.1--100~keV) luminosity can account for $\sim$20\% of the bolometric luminosity for both low-luminosity AGN, and for some of the high-luminosity AGN.


\section{Conclusions and Future Work}
In this work, we systematically investigate the coronal properties of 13 radio-quiet AGN at $z>1$ using quasi-simultaneous \NuSTAR\ and \XMM\ observations. This sample includes eight sources with newly acquired data. We conclude that:

\begin{enumerate}
\item The high-energy cutoff and coronal temperature are constrained at the 90\% confidence level for 10 and 9 targets, respectively, thanks to the deep \NuSTAR\ and \XMM\ exposures. For the remaining sources, the lack of constraints on $E_{\rm cut}$ and $kT_{\rm e}$ is most likely due to their low fluxes during the observations, which resulted in limited data quality.

\item In comparison, approximately 50\% of the low-$z$ AGN with \NuSTAR\ observations lack constrained cutoff energies, highlighting that high-$z$, luminous AGN provide a powerful opportunity to systematically investigate AGN coronal properties.

\item The mean cutoff energy of the high-$z$ AGN in our sample is $E_{\rm cut,mean,highz}$ = {101$\pm$21}~keV (1$\sigma$ uncertainties). For comparison, the mean cutoff energy of the local {\it Swift}/BAT AGN sample is $E_{\rm cut,mean,lowz}$ = { 204$\pm$31}~keV.

\item We find a potential correlation between cutoff energy and luminosity ($p$ = {0.025$_{-0.019}^{+0.057}$}) and black hole mass ($p$ = {0.044$_{-0.034}^{+0.10}$}), where sources with higher-luminosity and higher black hole mass have a lower cutoff energy. Such correlation is not measured between cutoff energy and Eddington ratio. Further confirming the correlations will require a large sample with good constraints on coronal properties. Hard X-ray mission concepts, like {\it HEX-P} \citep{Garcia2024}, will be essential for obtaining robust cutoff energy constraints for the bulk of the {\it Swift}/BAT AGN population \citep{Kammoun2024a}. 

\item The mean coronal temperature and optical depth of our PACHA sample are $kT_{\rm e}$ = {21.4$\pm$2.9} and $\tau$ = 4.8$\pm$0.2, respectively. 

\item Under the hybrid coronal framework, the systematically low coronal temperatures of the high-$z$, luminous AGN suggest that the electron population in the corona cannot be purely thermal. Low coronal temperatures may indicate either a large fraction of non-thermal electrons in the corona and/or highly efficient Compton cooling. Clear predictions of a large fraction of non-thermal electrons are the emergence of an excess at hard X-rays at a few hundred keV and a broad annihilation line at 511~keV. Both can be tested by future MeV telescopes, e.g., {\it COSI} \citep{Tomsick2024}. 

\item The low coronal temperatures can also be explained by a purely thermal corona based on radiation MHD simulations, in which magnetic buoyancy transports accretion power into the corona. As the accretion rate increases, the fraction of power released in the corona declines, leading to systematically lower coronal temperatures.

\item X-ray (0.1--100~keV) emissions could contribute up to $\sim$20\% of the entire bolometric luminosity in some high-$z$, luminous AGN.

\item Future work incorporating hybrid frameworks with more realistic geometries is required to robustly constrain the physical structure and energy dissipation mechanisms of the corona. 

\item We present potential correlations between coronal properties and luminosity and black hole mass, which could serve as important inputs for cosmic X-ray background (CXB) population synthesis models \citep[e.g.,][]{gilli07,Ueda14,Buchner2015,Tasnim_Ananna_2019}, enabling tighter constraints on the contribution of luminous AGN to the integrated CXB and offering better constraints on the cosmic SMBH accretion history.

\item Current coronal models assume a homogeneous corona, which can be improved in future works as the corona is found to be significantly inhomogeneous from AGN variability studies \citep[e.g.,][]{Zhao2025} and from the GRHMD simulations \citep[e.g.,][]{Jiang2019} and the radiative particle-in-cell simulations \citep[e.g.,][]{Grovselj2026}.

\item{{Another four high-$z$ AGN have been awarded 500 ks of NuSTAR and 125 ks of XMM-Newton observing time in NuSTAR Cycle 12 (PI: Zhao, proposal ID 12110) and will be included in future papers of the PACHA series.}}
\end{enumerate}

\section{acknowledgments}
{We thank the anonymous referee for their helpful comments.} X.Z. acknowledges NASA funding under contract numbers 80NSSC23K1648 and 80NSSC24K1031. X.Z. thank Karl Foster and Murray Brightman for their help in designing the observation plan and scheduling the observations. X.Z. appreciates Brian Grefenstette for the helpful discussion about \NuSTAR\ data reduction. X.Z. appreciates Warren Brown and Gilbert Esquerdo for scheduling and conducting the FAST observation. X.Z. appreciates George Chartas for his helpful discussion on the magnification of the lens-system J0913+5259. {X.Z. appreciates Daniel Groselj for the helpful discussion on non-thermal corona physics.} ChatGPT (OpenAI) has been used to generate Fig.~\ref{fig:illustration}.

L.C. acknowledges support from NASA ATP grant 80NSSC24K1230. E.B. acknowledges financial support from INAF under the Large Grant 2022 “The metal circle: a new sharp view of the baryon cycle up to Cosmic Dawn with the latest generation IFU facilities” and from the “Ricerca Fondamentale 2024” program under the GO grant “A JWST/MIRI MIRACLE: Mid-IR Activity of Circumnuclear Line Emission” and the mini-grant 1.05.24.07.01.

This work has made use of data from the \NuSTAR\ mission, a project led by the California Institute of Technology, managed by the Jet Propulsion Laboratory, and funded by the National Aeronautics and Space Administration. 
This research has made use of the \NuSTAR\ Data Analysis Software (NuSTARDAS) jointly developed by the ASI Science Data Center (ASDC, Italy) and the California Institute of Technology (USA). 

This research has made use of data and software provided by the High Energy Astrophysics Science Archive Research Center (HEASARC), which is a service of the Astrophysics Science Division at NASA/GSFC and the High Energy Astrophysics Division of the Smithsonian Astrophysical Observatory. 

This work is based on observations obtained with \XMM, an ESA science mission with instruments and contributions directly funded by ESA Member States and NASA. 

This work makes use of the data from SDSS. Funding for the Sloan Digital Sky Survey has been provided by the Alfred P. Sloan Foundation, the U.S. Department of Energy Office of Science, and the Participating Institutions. 

This publication makes use of data products from the Two Micron All Sky Survey (2MASS), which is a joint project of the University of Massachusetts and the Infrared Processing and Analysis Center/California Institute of Technology, funded by the National Aeronautics and Space Administration and the National Science Foundation.

This work is partly based on the data from WISE, which is a joint project of the University of California, Los Angeles, and the Jet Propulsion Laboratory/California Institute of Technology.

The DESI Legacy Imaging Surveys consist of three individual and complementary projects: the Dark Energy Camera Legacy Survey (DECaLS), the Beijing-Arizona Sky Survey (BASS), and the Mayall z-band Legacy Survey (MzLS). DECaLS, BASS and MzLS together include data obtained, respectively, at the Blanco telescope, Cerro Tololo Inter-American Observatory, NSF’s NOIRLab; the Bok telescope, Steward Observatory, University of Arizona; and the Mayall telescope, Kitt Peak National Observatory, NOIRLab. NOIRLab is operated by the Association of Universities for Research in Astronomy (AURA) under a cooperative agreement with the National Science Foundation. Pipeline processing and analyses of the data were supported by NOIRLab and the Lawrence Berkeley National Laboratory (LBNL). Legacy Surveys was supported by: the Director, Office of Science, Office of High Energy Physics of the U.S. Department of Energy; the National Energy Research Scientific Computing Center, a DOE Office of Science User Facility; the U.S. National Science Foundation, Division of Astronomical Sciences; the National Astronomical Observatories of China, the Chinese Academy of Sciences and the Chinese National Natural Science Foundation. LBNL is managed by the Regents of the University of California under contract to the U.S. Department of Energy. 

The National Radio Astronomy Observatory (NRAO) is a facility of the National Science Foundation operated under cooperative agreement by Associated Universities, Inc. This paper makes use of observations from the Karl G. Jansky Very Large Array (VLA) of the NRAO.


\appendix
\renewcommand{\thesection}{APPENDIX~\Alph{section}}

\begingroup
\renewcommand*{\arraystretch}{1.2}
\begin{table*}
\begin{center}
\caption{Summary of \NuSTAR\ and \XMM\ observations analyzed in this work. \tablenotemark{a}{{The reported exposures are the cleaned exposure times of FPMA and FPMB for NuSTAR, and of MOS1, MOS2, and pn for XMM-Newton, respectively, after removing the high-background intervals.}}
\tablenotemark{b}{{ The reported \NuSTAR\ count rates are those of the FPMA and FPMB modules in the 3--24\,keV range, respectively. The reported \XMM\ count rates are those of the MOS1, MOS2, and pn modules in the 0.3--10\,keV range, respectively.}}}\label{Tab:obs}
  \begin{tabular}{cccccc}
       \hline
       \hline
    Target&Telescope&Sequence&Start Time&Exposure\tablenotemark{a}&Net Count Rate\tablenotemark{b}\\ 
    &&ObsID&(UTC)&(ks)&($10^{-2}$ counts s$^{-1}$)\\
    \hline
           \hline
    WISEA J094835.95+432302.6&\NuSTAR&60861002002&2022-10-31 T09:26:09&80/80&0.42/0.51\\
    &\NuSTAR&60861002004&2022-11-02 T17:41:09&86/85&0.49/0.45\\
    &\XMM&0913070401&2022-10-31 T11:08:44&40/43/32&3.1/3.4/12\\
    \hline
    WISEA J122527.39+223512.9&\NuSTAR&60861001002&2022-12-04T08:11:09&141/139&0.33/0.23\\
    &\XMM&0913070101&2022-12-06T08:45:38&37/43/18&2.0/2.4/8.9\\
    \hline
        WISEA J163428.99+703132.4&\NuSTAR&60961001002&2023-06-28 T13:36:09&80/80&0.20/0.16\\
    &\XMM&0930830101&2023-06-28 T20:28:29&20/22/17&11.7/14.2/54\\
           \hline
       SDSS J145453.53+032456.8&\NuSTAR&60401018002&2019-01-12 T23:51:09&116/116&0.78/0.79\\
       &\NuSTAR&60401018004&2019-01-18 T05:56:09&69/69&0.48/0.47\\
    &\XMM&0830480201&2019-01-14 T11:25:32&38/38/25&5.7/7.1/26\\          
      \hline
    WISEA J091301.04+525928.7&\NuSTAR&60861003002&2023-04-20 T16:16:09&55/54&3.3/3.2\\
    &\NuSTAR&60861003002&2023-04-27 T09:01:09&97/96&3.0/2.8\\
    &\XMM&0913070201&2023-04-20 T23:53:34&17/24/12&21/21/81\\
    \hline
    SDSS J092115.47+285444.3 &\NuSTAR&60961002002&2023-10-18 T20:01:09&117/116&0.17/0.14\\
    &\XMM&0930830201&2023-10-19 T12:54:13&30/31/23&7.5/8.6/31\\
           \hline
    WISEA J145848.63+525450.3&\NuSTAR&60961003002&2023-11-28 T08:46:09&142/144&0.25/0.15\\
    &\XMM&0930830301&2023-11-28 T09:13:35&30/35/19&2.1/2.3/8.9\\
           \hline    
    WISEA J123343.47+161214.9&\NuSTAR&60961004002&2023-12-02 T08:56:09&156/155&0.28/0.23\\
    &\XMM&0930830401&2023-12-04 T08:05:47&48/57/31&1.7/2.0/7.5\\
        \hline
    WISEA J125948.78+342322.5&\NuSTAR&60861004002&2023-05-30 T07:26:09&69/69&0.55/0.44\\
    &\NuSTAR&60861004004&2023-06-06 T11:21:09&84/83&0.42/0.36\\
    &\XMM&0913070301&2023-05-31 T01:49:41&40/41/33&3.2/3.6/15\\
    \hline
       \hline  
       2MASS J1614346+470420&\NuSTAR&60301021002&2017-08-25 T00:11:09&100/99&0.77/0.80\\
       &\NuSTAR&60301021004&2017-10-07 T04:21:09&47/46&0.67/0.78\\
    &\XMM&0803910101&2017-08-05 T17:53:48&87/87/69&4.4/6.0/24\\
    \hline
       B1422+231&\NuSTAR&60301020002&2017-12-30 T21:06:09&98/97&1.1/1.0\\
    &\XMM&0795640101&2017-12-29 T07:06:24&35/35/26&7.1/6.9/23\\
    \hline
       APM 08279+5255&\NuSTAR&60401017002&2019-04-19 T21:41:09&93/93&0.42/0.39\\
       &\NuSTAR&60401017004&2019-04-22 T12:21:09&60/59&0.41/0.45\\
    &\XMM&0830480301&2019-04-23 T22:34:13&29/29/25&2.0/2.3/7.2\\   
    \hline
          PG 1247+267&\NuSTAR&60001030002&2014-12-04 T08:51:07&61/72&0.6/0.7\\
    &\XMM&0143150201&2003-06-18 T09:20:58&35/35/26&5.2/4.8/20\\

       \hline
       \hline
\end{tabular}
\par
\vspace{.3cm}
\end{center}
\end{table*}
\endgroup

\section{Data Reduction}\label{sec:DR}
The observations used in this work are summarized in Table~\ref{Tab:obs}. The \NuSTAR\ observations of a few targets were divided into two observations separated by a few days due to interruptions by Targets of Opportunity (ToOs).

\subsection{\NuSTAR\ Data Reduction}
We reduced the \NuSTAR\ data of all 13 targets using HEASoft v.6.32.1 with the updated calibration and response files CALDB v.20240325. The level 1 raw data were calibrated, cleaned, and screened by running the \texttt{nupipeline} script. To remove time intervals at South Atlantic Anomaly (SAA) when the background event rates are higher, we used the options \texttt{saacalc=3}, \texttt{saamode=OPTIMIZED}, and \texttt{tentacle=yes}, except for ObsID = 60861002004, whose background is better cleaned with \texttt{saacalc=1}. The solar flares affected the \NuSTAR\ observations of four targets: ObsIDs 60961004002, 60961001002, 60961003002, 60861004002, and 60861004004. For these observations, we removed time intervals when their count rates across the field-of-view (FoV) were $\ge$2~$\sigma$ higher than the average count rate of the entire observation. We applied the good time intervals using \texttt{nupipeline} option {\tt usrgtifile}. 

As the \NuSTAR\ background is spatially uneven across the FoV, we modeled the \NuSTAR\ background using the {\tt nuskybgd} tool\footnote{\url{https://github.com/NuSTAR/nuskybgd}} \citep{Wik_2014}, which is widely used in the background analysis of \NuSTAR\ observations of extended, faint targets and extragalactic surveys. We followed standard protocols when applying {\tt nuskybgd} to our observations \citep[e.g., following][]{Wik_2014,zhao2024b}. We verified the quality of the simulated backgrounds following \citet{Zhao2021} before we applied them to the spectral analysis. In the end, the 3--10~keV source fluxes measured in \NuSTAR\ are in good agreement with what was measured by \XMM, indicating that the \NuSTAR\ backgrounds are well modeled. Unexpectedly bright emission (likely caused by solar flares) is found on FPMA detector 2 and 3 of PG 1247+267 (detector 1 where the source lies is not affected). This could not be well treated with {\tt nuskybgd}, so, we extracted its background spectrum using a circular region (with a radius of 80\arcsec) close to the source region. 

The source spectra of the eight targets were extracted from circular regions with radii ranging from 35\arcsec\ to 50\arcsec\ (the radii can be different for FPMA and FPMB), corresponding to encircled energy fractions (EEFs) of 50\%--70\% at 10~keV. The radii were determined by maximizing both the S/N and the number of net counts of the source spectra following the method in \citet{Zappacosta_2018}. Typically, the spectra of brighter sources were extracted using regions with larger radii. The background spectra were extracted using the same regions as were used in extracting the source spectra.

\subsection{\XMM\ Data Reduction}\label{sec:XMM}
The \XMM\ observations of the 13 targets were taken contemporaneously with the \NuSTAR\ observations (Table~\ref{Tab:obs}). We reduced the \XMM\ observations following the standard process \citep[e.g.,][]{zhao2024b}. 

The \XMM\ data were reduced using the Science Analysis System \citep[SAS;][]{SAS} version 21.0.0. Due to the strong background flares, 13--67\% of the \XMM\ exposures were lost. The effective exposures used for spectral analysis of the eight targets are reported in Table~\ref{Tab:obs}. The source spectra were extracted from a radius of 20\arcsec\ circular region for each target, corresponding to 77\% of the EEF at 1.5\,keV. The background spectra were extracted from radius of 40\arcsec\ circle regions located near the targets but avoiding nearby sources. We analyzed the \XMM\ spectra between 0.5 and 10~keV from all three cameras, i.e., MOS1, MOS2, and pn.

It may be worth noting that a flux difference of $\sim$10\% has been reported between \XMM\ and \NuSTAR\ \citep{Madsen2017}. In addition, slight differences in the spectral shapes measured by the two observatories have been identified\footnote{https://xmmweb.esac.esa.int/docs/documents/CAL-TN-0230-1-3.pdf}, which can lead to systematic offsets in the derived photon indices of up to $\Delta\Gamma \lesssim 0.1$ \citep{Kang2023}. These calibration differences may therefore introduce modest systematic effects on the inferred high-energy cutoff and coronal temperature. However, the impact is expected to be limited, as the statistical uncertainties on $E_{\rm cut}$ and $kT_{\rm e}$ in our analysis are substantially larger than the systematic uncertainties associated with the cross-calibration of the two telescopes.

\section{Spectral Analysis Details}\label{sec:spectral_analysis}
The best-fit values of the spectra of each source using all four models are reported in Table~\ref{Tab:all_source}. Their best-fit spectra are plotted in Fig.~\ref{fig:spectrum1} to Fig.~\ref{fig:spectrum3}.

{WISEA J091301.04+525928.7}: 
In its M CP modeling, we fixed the ionization parameter at log($\xi$) = 2 as the contribution of the reflection component to the entire spectrum is negligible. $C_{N1/X}$ = and where $C_{N1/X}$ is the cross calibration between the first \NuSTAR\ exposure and \XMM. $C_{N1/X}>$1 suggests that \NuSTAR\ flux is higher than the \XMM\ flux. $C_{N2/X}$ is the cross calibration between the second \NuSTAR\ observation (if there is any) and \XMM. 

{SDSS J092115.47+285444.3}:
The source spectra exhibit a prominent iron K$\alpha$ emission line at a rest-frame energy of $\sim$6.4~keV. The initial best fit with M CP, assuming solar iron abundance (A$_{\rm Fe}$ = 1), yields a $C$-stat/dof of 492/376, but fails to adequately model the line. Allowing the iron abundance to vary significantly improves the fit, though only lower limits on A$_{\rm Fe}$ are obtained. {The best-fit A$_{\rm Fe}$ are $>$9 and $>$7.3 for model {\tt PEX} and {\tt CP}, respectively.}

These results indicate that the reflection component produces a strong iron K$\alpha$ line but only a moderate Compton hump at 10--30~keV. This likely suggests that the reflection originates from material with a lower density than assumed in the $xillverCp$ model. Since the minimum density permitted in these models is 10$^{15}$~cm$^{-3}$, an unusually high A$_{\rm Fe}$ is required to compensate for the high-density assumption. Nevertheless, we cannot rule out the possibility that the accretion disk in this source is genuinely iron-enriched, with abundances exceeding the Solar value by at least a factor of a few.


{SDSS J145453.53+032456.8}:
The flux of J1454+0324 during the second \NuSTAR\ observation is $\sim$35\% lower than the flux during the first \NuSTAR\ observation.

{WISEA J163428.99+703132.4}:
J1634+70313 presents an excess in soft X-rays ($\le$1~keV); therefore, we added a {\tt zbbody} when modeling the spectra. The reduced $\chi^2$ is improved from 457/369 to 430/367 by adding the {\tt zbbody} component. {The best-fit temperature of the {\tt zbbody} component are 0.24$_{-0.05}^{+0.04}$, 0.26$_{-0.04}^{+0.03}$, 0.22$_{-u}^{+u}$, and 0.24$_{-0.08}^{+0.06}$ keV for model {\tt powerlw}, {\tt cutoff}, {\tt PEX}, and {\tt CP}, respectively.}

{WISEA J094835.95+432302.6}:
In its {M CP} modeling, we fixed the ionization parameter at log($\xi$) = 2 as the contribution of the reflection component to the entire spectrum is negligible. {Potential ultra-fast inflow and outflow signals have been detected in the \XMM\ spectrum of the source (Xu et. al., in prep.)}
 
{PG 1247+267}:
The \NuSTAR\ and \XMM\ observations were taken at different epochs, separated by 11.5 years, leading to a flux variability of a factor of $\sim$2.

{APM 08279+5255}:
\citet{Bertola2022} measured a cutoff energy $E_{\rm cut}$ = 99$_{-35}^{+91}$~keV, which is consistent with our measurements. {The source has a measured intrinsic absorption of 15$\pm$2, 8$\pm$2, 8$\pm$1, and 13$_{-5}^{+2}$ $\times$10$^{22}$~cm$^{-2}$ for the powerlw, cutoff, PEX, and CP models, respectively.}

{B1422+231}:
\citet{Lanzuisi2019} measured a cutoff energy of $E_{\rm cut}$ = 66$_{-12}^{+17}$~keV, which is consistent with our measurements.

{2MASS J1614346+470420}:
\citet{Lanzuisi2019} measured a cutoff energy of $E_{\rm cut}$ = 106$_{-37}^{+102}$~keV, which is consistent with our measurements.

{PG 1247+267}:
\citet{Lanzuisi2016} measured a cutoff energy of $E_{\rm cut}$ = 89$_{-34}^{+112}$~keV, which is consistent with our measurements.

\section{Luminosity vs Black Hole Mass}\label{sec:BH_lum}
\begin{figure}[t] 
\centering
\includegraphics[width=.49\textwidth]{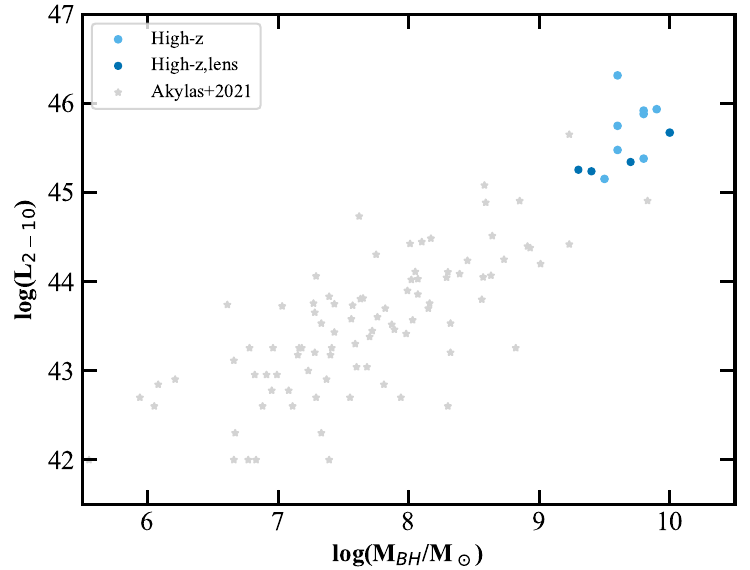}
\caption{2--10~keV luminosity as a function of the black hole mass of the sources from the PACHA sample (blue) and the \citet{Akylas2021} sample (grey).}
\label{fig:BH_lum}
\end{figure}

\begin{figure*}[t] 
\centering
\includegraphics[width=.87\textwidth]{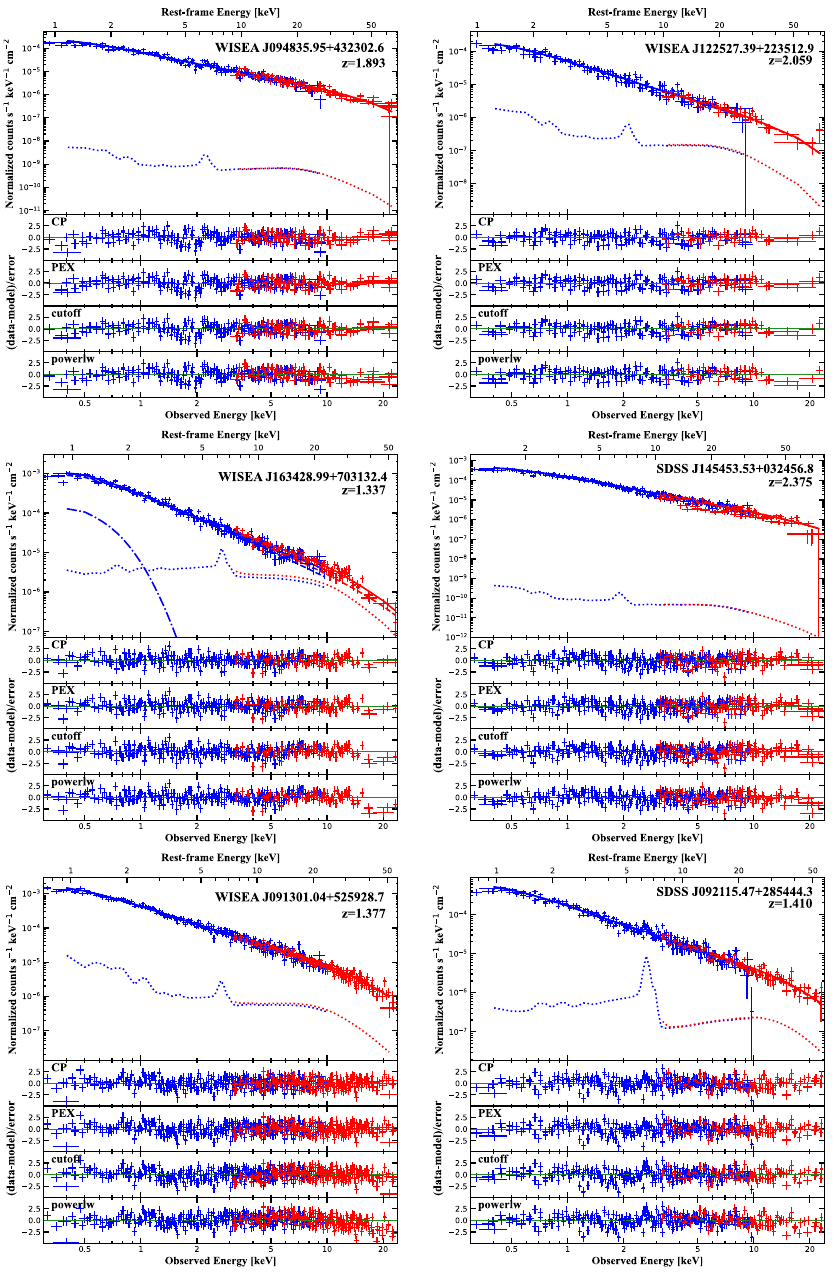}
\caption{Normalized best-fit spectra of the PACHA targets with M CP and the residuals from all four models. NuSTAR and XMM-Newton data are plotted in red and blue, respectively. The total model prediction, line-of-sight continuum, reflection, and blackbody components are plotted in solid, dashed, dotted, dash-dot lines, respectively.}
\label{fig:spectrum1}
\end{figure*}

\begin{figure*}[t] 
\centering
\includegraphics[width=.88\textwidth]{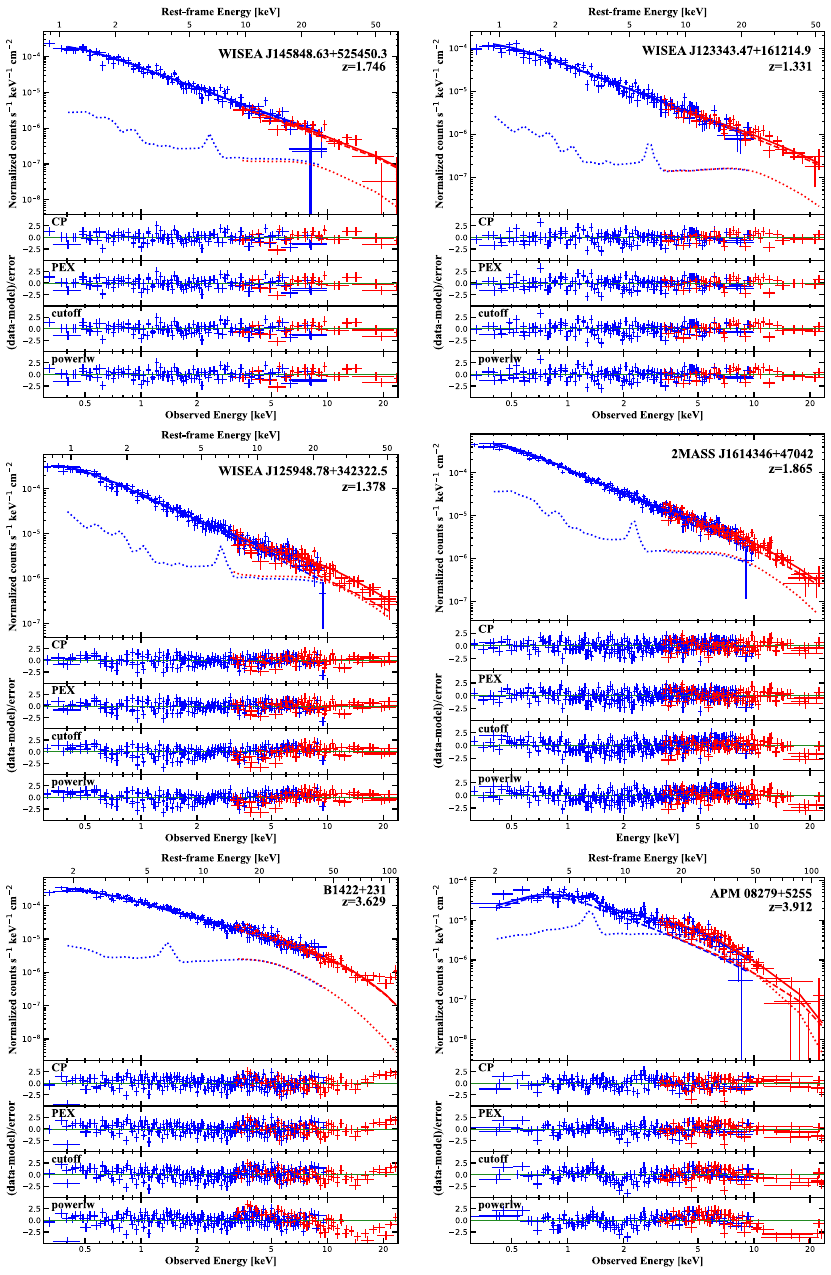}
\caption{Figure~\ref{fig:spectrum1} continued.}
\label{fig:spectrum2}
\end{figure*}

\begin{figure}[t] 
\centering
\includegraphics[width=.48\textwidth]{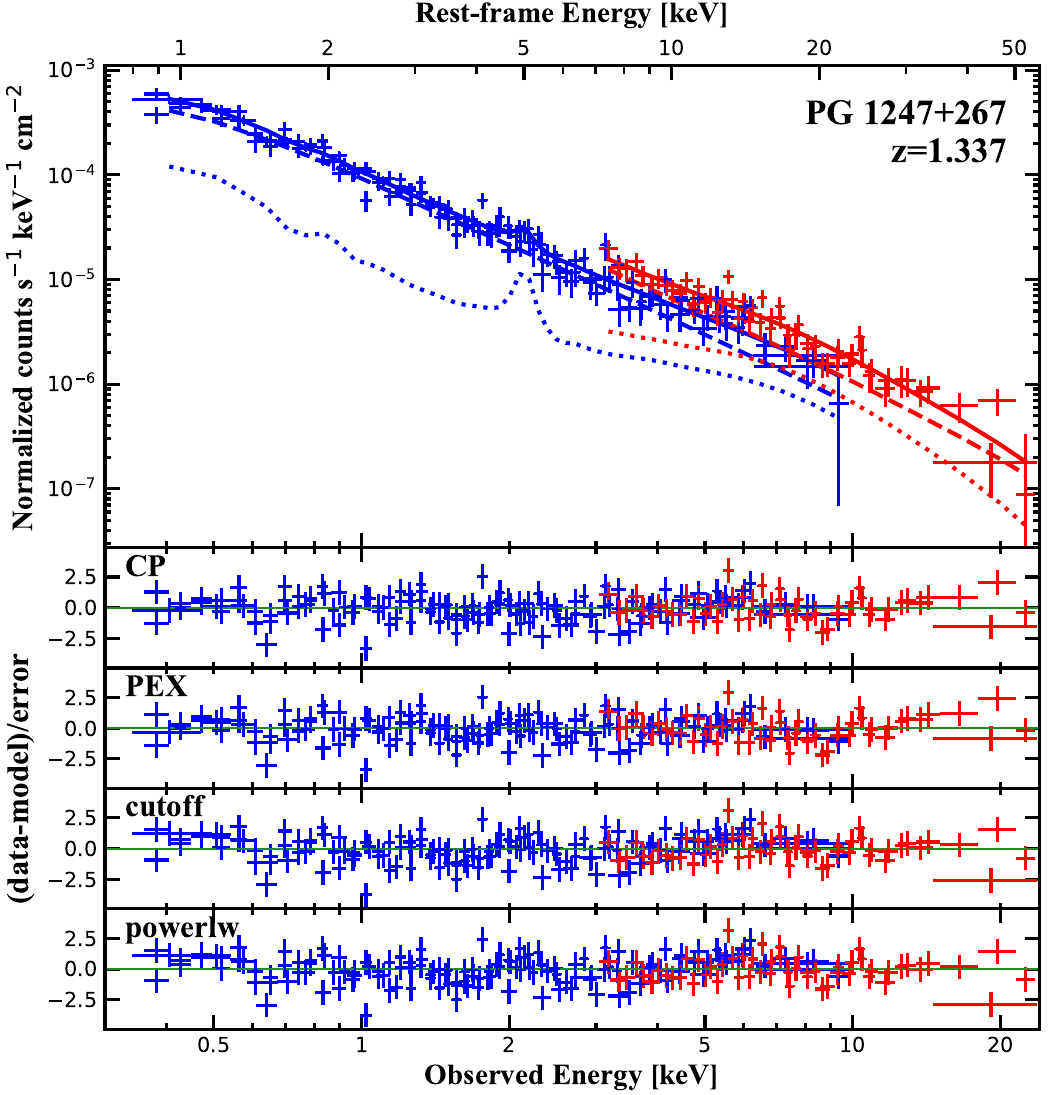}
\caption{Figure~\ref{fig:spectrum1} continued.}
\label{fig:spectrum3}
\end{figure}

There is a clear correlation between the 2--10~keV luminosity and black hole mass of the sources from both the PACHA sample and the \citet{Akylas2021} sample (Fig.~\ref{fig:BH_lum}), suggesting a selection bias.

\section{Coronal Size Derived using Micro-lensing Method} \label{sec:mirco}
The coronal size derived using the micro-lensing method \citep{Chartas2016} is reported in Table~\ref{Table:CS_micro}.

\begingroup
\renewcommand*{\arraystretch}{1.2}
\begin{table}
\begin{center}
\caption{Global coronal size of a few AGN measured using microlensing methods. The black hole mass, half-light radius (HLR) of the corona, and coronal size (CS) of the sources are adopted from \citet{Chartas2016}.}\label{Table:CS_micro}
  \begin{tabular}{cccccccc}
       \hline
       \hline
    Source&$z$&log(M$\rm_{\rm BH}$)&HLR&CS\\ 
    &&(M$_{\odot}$)&(10$^{-4}$~pc)&($r\rm _g$)\\       
        \hline
RX J1131--1231&0.658&7.8$_{-0.6}^{+0.2}$&0.3$_{-0.1}^{+0.5}$&14$_{-6}^{+13}$\\   
Q J0158-4325&1.29&8.2$_{-0.1}^{+0.1}$&0.7$_{-0.5}^{+1.0}$&10$_{-5}^{+9}$\\    
SDSS 0924+0219&1.524&8.4$_{-0.5}^{+0.3}$&0.8$_{-0.6}^{+3.4}$&10$_{-6}^{+15}$\\
PG 1115+080&1.722&8.7$_{-0.3}^{+0.1}$&13$_{-11}^{+40}$&88$_{-50}^{+116}$\\            
HE 0435-1223&1.689&8.7$_{-0.4}^{+0.2}$&$<$5&$<$59\\    
HE 1104-1805&2.32&8.8$_{-0.3}^{+0.1}$&$<$7&$<$46\\  
Q 2237+0305&1.695&9.1$_{-0.4}^{+0.2}$&21$_{-13}^{+32}$&44$_{-19}^{+36}$\\
        \hline
         \hline     
\end{tabular}
\end{center}
\end{table}
\endgroup

\begin{table*}[t]
\renewcommand*{\arraystretch}{1.07}
\centering
\footnotesize
\caption{Best-Fit Results. $\Gamma$ is the photon index. $E_{\rm cut}$ is the cutoff energy in keV. $R$ is the reflection scaling factor. $kT_{\rm e}$ is the electron temperature of the corona in keV. log$\xi$ is the Ionization parameter. F$_{0.5-2}$ and F$_{2-10}$ are the \XMM\ measured flux between 0.5--2~keV and 2--10~keV in $10^{-13}$ erg cm$^{-2}$ s$^{-1}$, respectively. L$_{2-10}$ \XMM\ measured intrinsic (delensed) luminosity between 2--10\,keV in $10^{45}$ erg s$^{-1}$. $^f$ is when the parameter is fixed at a given value when fitting the spectra.}\label{Tab:all_source}
\vspace{-0.4cm}

  \begin{tabular}{ccccccccccccccccc}
       \hline
       \hline       
       Target&Model&$C$/dof&$\Gamma$&$E_{\rm cut}$&$R$&$kT_{\rm e}$&log$\xi$&F$_{0.5-2}$&F$_{2-10}$&L$_{2-10}$\\
    \hline
	J0948+4323
    &powerlw&493/452&1.66$\pm$0.03&-&-&-&-&1.51$\pm$0.03&3.0$\pm$0.1&5.5$\pm$0.1\\
    &cutoff&469/451&1.55$\pm$0.05&61$_{-16}^{+33}$&-&-&-&1.51$\pm$0.03&2.9$\pm$0.1&5.6$\pm$0.1\\
    &PEX&469/450&1.56$\pm$0.04&59$_{-18}^{+20}$&0.05$_{-u}^{+0.27}$&-&-&1.51$\pm$0.03&2.9$\pm0.1$&5.6$\pm$0.1\\
    &CP&484/451&1.69$\pm$0.02&-&-&13.5$_{-2.5}^{+4.9}$&2$^f$&1.50$\pm$0.03&3.0$_{-0.1}^{+0.2}$&5.5$\pm$0.1\\    
        \hline
    J1225+2235
    &powerlw&332/336&1.79$\pm$0.04&-&-&-&-&1.14$_{-0.04}^{+0.03}$&1.9$_{-0.2}^{+0.1}$&4.8$_{-0.1}^{+0.2}$\\
    
    &cutoff&327/335&1.72$\pm$0.07&$89_{-39}^{+230}$&-&-&-&1.14$\pm$0.03&1.8$\pm$0.01&4.9$\pm$0.1\\
    
    &PEX&325/334&$1.76_{-0.08}^{+0.09}$&$70_{-23}^{+100}$&$0.3_{-u}^{+0.4}$&-&-&$1.13_{-0.04}^{+0.03}$&1.9$\pm$0.1&4.7$\pm$0.2
        \\
    &CP&326/334&$1.81_{-0.04}^{+0.05}$&-&-&$15_{-4}^{+14}$&$2^f$&$1.13_{-0.04}^{+0.03}$&1.9$\pm$0.1&4.7$\pm$0.1
        \\
        \hline

        J1634+7031
    &powerlw
    &430/367
    &2.01$\pm$ 0.04
    &-
    &-
    &-
    &-
    &6.7$\pm$0.1
    &8.3$\pm$0.3
    &9.3$\pm$0.2
    \\
    
        &cutoff
        &423/366
        &$1.91_{-0.08}^{+0.07}$
        &$97_{-39}^{+170}$
        &-
        &-
        &-
        &6.7$\pm$0.1
        &8.4$\pm$0.3
        &9.3$\pm$0.2
        
        \\
        
        &PEX
        &398/365
        &$2.11_{-0.10}^{+0.11}$
        &$66_{-18}^{+28}$
        &$0.99_{-0.44}^{+0.63}$
        &-
        &-
        &6.7$\pm$0.1
        &8.5$\pm$0.3
        &8.5$\pm$0.2
        \\
        
        &CP
        &399/365
        &$2.09_{-0.04}^{+0.06}$
        &-
        &-
        &$15.4_{-3.2}^{+5.3}$
        &$<1.4$
        &6.7$\pm$0.1
        &8.5$\pm$0.3
        &8.7$\pm$0.3
    \\
	\hline
        J1454+0324
    &powerlw&533/474&1.86$\pm$0.02&-&-&-&-&3.31$\pm$0.05&5.2$_{-0.1}^{+0.2}$&20.6$\pm$0.3\\
	&cutoff&522/473&1.81$\pm$0.03&140$_{-50}^{+160}$&-&-&-&3.32$\pm$0.05&5.2$\pm$0.2&20.7$\pm$0.4\\
	&PEX&522/472&1.81$\pm$0.04&138$_{-50}^{+160}$&0.02$_{-u}^{+0.16}$&-&-&3.32$\pm$0.05&5.2$\pm0.2$&20.7$\pm$0.3\\
	&CP&523/472&1.86$\pm$0.02&-&-&27$_{-7.6}^{+36}$&2$^f$&3.31$\pm$0.06&5.3$_{-0.2}^{+0.1}$&20.6$_{-0.3}^{+0.4}$\\
       \hline
       J0913+5259
    & 
    powerlw& 
    691/516& 
    $1.83_{-0.02}^{+0.01}$& 
    -& 
    -& 
    -& 
    -& 
    9.9$\pm$0.1& 
    15.4$\pm$0.4& 
    1.79$\pm$0.03 
    \\
    
        & 
        cutoff& 
        592/515& 
        $1.72_{-0.03}^{+0.02}$& 
        $64_{-10}^{+13}$& 
        -& 
        -& 
        -& 
        10.3$\pm$0.1& 
        15.6$\pm$0.4&
        1.84$_{-0.02}^{+0.03}$
        \\

        & 
        PEX& 
        571/514& 
        1.72$_{-0.02}^{+0.03}$&
        42$_{-5}^{+7}$& 
        0.40$_{-0.15}^{+0.17}$& 
        -& 
        -& 
        9.8$\pm$0.1& 
        16.0$_{-0.4}^{+0.5}$& 
        1.77$\pm$0.02
        \\
            
        & 
        CP& 
        563/514& 
        1.81$\pm$0.02& 
        -& 
        -& 
        11.4$_{-0.8}^{+1.0}$& 
        2$^f$& 
        9.8$\pm$0.1& 
        16.2$_{-0.4}^{+0.3}$& 
        1.80$_{-0.02}^{+0.03}$
        \\
    \hline 
 
        J0921+2854
    &powerlw&502/379&1.66$\pm$0.02&-&-&-&-&3.84$_{-0.05}^{+0.06}$&7.7$_{-0.2}^{+0.3}$&1.83$\pm$0.03\\
	&cutoff&480/378&1.58$\pm$0.03&82$_{-21}^{+43}$&-&-&-&3.85$_{-0.06}^{+0.05}$&7.7$_{-0.3}^{+0.2}$&1.85$\pm$0.03\\
	&PEX&468/377&1.57$\pm$0.03&71$_{-16}^{+27}$&0.24$_{-0.14}^{+0.13}$&-&-&3.83$\pm$0.06&7.6$_{-0.2}^{+0.3}$&1.83$\pm$0.03\\
	&CP&482/375&1.70$\pm$0.02&-&-&12.5$_{-1.6}^{+2.4}$&$<$1.5&3.80$\pm$0.06&7.9$_{-0.2}^{+0.3}$&1.79$_{-0.03}^{+0.04}$\\
       \hline
       
    J1458+5254
    &powerlw&361/331
    &1.93$_{-0.04}^{+0.05}$&-&-&-&-&1.13$_{-0.04}^{+0.03}$&1.5$\pm$0.1&3.0$\pm$0.1\\
    &cutoff&361/330
    &{1.93$_{-0.07}^{+0.03}$}&$>$80&-&-&-&1.13$_{-0.04}^{+0.03}$&1.5$\pm$0.1&3.0$\pm$0.1\\
    &PEX&361/329
    &1.94$\pm$0.08&$>$65&0.1$_{-u}^{+0.5}$&-&-&1.12$_{-0.03}^{+0.04}$&1.5$_{-0.1}^{+0.2}$&3.0$\pm$0.1\\
    &CP&360/329
    &1.94$_{-0.05}^{+0.03}$&-&-&$>$14&2$^f$&1.12$_{-0.03}^{+0.04}$&1.5$_{-0.1}^{+0.2}$&3.0$\pm$0.1\\
        \hline
        
    J1233+1612
    &powerlw&401/344
    &1.70$\pm$0.04&-&-&-&-&0.91$\pm$0.02&1.7$\pm$0.1&1.48$_{-0.04}^{+0.05}$\\
    &cutoff&401/343
    &{1.70$_{-0.05}^{+0.02}$}&{$>$140}&-&-&-&0.91$\pm$0.02&1.7$\pm$0.1&1.48$_{-0.04}^{+0.05}$\\
    &PEX&397/342
    &1.74$\pm$0.06&$>$49&0.56$_{-0.45}^{+0.60}$&-&-&0.91$_{-0.02}^{+0.03}$&1.8$\pm$0.1&1.41$\pm$0.05\\
    &CP&399/343
    &1.74$_{-0.05}^{+0.04}$&-&-&$>$13&2$^f$&0.91$_{-0.03}^{+0.02}$&1.8$\pm$0.1&1.44$\pm$0.05\\
        \hline    
      J1259+3423
    &powerlw&468/451&1.88$\pm$0.03&-&-&-&-&1.79$_{-0.04}^{+0.03}$&2.6$_{-0.2}^{+0.1}$&2.8$\pm$0.1\\
    &cutoff&468/450&1.87$_{-0.02}^{+0.03}$&{$>$700}&-&-&-&1.79$\pm$0.01&2.5$\pm$0.1&2.8$\pm$0.1\\
    &PEX&421/448&2.02$_{-0.05}^{+0.06}$&$>$70&1.7$_{-0.6}^{+0.8}$&-&-&1.75$\pm$0.04&2.8$\pm$0.1&2.4$\pm$0.1\\
    &CP&424/448&1.96$_{-0.03}^{+0.10}$&-&-&$>$20&$<$2.1&1.75$_{-0.04}^{+0.03}$&2.8$\pm$0.1&2.4$\pm$0.1\\
        \hline
 J1614+4704
    &powerlw&526/500&1.86$\pm$0.02&-&-&-&-&2.87$_{-0.03}^{+0.04}$&4.2$\pm$0.1&9.3$\pm$0.1\\
    &cutoff&{526/499}&{1.86$\pm$0.02}&{$>$600}&-&-&-&2.87$\pm$0.03&4.2$\pm$0.1&9.3$\pm$0.1\\
    &PEX&466/498&1.94$\pm$0.03&97$_{-27}^{+57}$&1.0$\pm$0.3&-&-&2.83$_{-0.03}^{+0.04}$&4.5$\pm$0.1&8.5$\pm$0.1\\
    &CP&463/497&1.90$_{-0.03}^{+0.04}$&-&-&21$_{-4}^{+16}$&2.1$\pm$0.4&2.83$\pm$0.03&4.5$\pm$0.1&8.1$\pm$0.2\\
        \hline

    B1422+231
    &powerlw&585/379&1.63$\pm$0.02&-&-&-&-&2.90$\pm$0.05&6.3$_{-0.2}^{+0.3}$&2.32$\pm$0.04\\  
    &cutoff&464/378&1.42$\pm$0.04&53$_{-8}^{+10}$&-&-&-&2.92$\pm$0.05&6.0$_{-0.1}^{+0.2}$&2.32$\pm$0.04\\   
    &PEX&444/377&1.57$\pm$0.06&58$_{-9}^{+12}$&0.44$_{-0.17}^{+0.21}$&-&-&2.88$\pm$0.05&6.2$\pm$0.2&2.27$\pm$0.04\\  
    &CP&468/377&1.69$_{-0.03}^{+0.04}$&-&-&15.6$_{-1.8}^{+2.2}$&2$^f$&2.85$\pm$0.05&6.4$\pm$0.2&2.21$\pm$0.04\\ 
        \hline
     APM 08279
    &powerlw&592/434&1.87$\pm$0.07&-&-&-&-&0.96$\pm$0.03&2.3$_{-0.1}^{+0.2}$&7.5$_{-0.5}^{+0.8}$\\
    &cutoff&490/433&1.16$_{-0.14}^{+0.13}$&28$_{-4}^{+7}$&-&-&-&0.92$_{-0.04}^{+0.03}$&2.5$\pm$0.1&5.3$_{-0.3}^{+0.5}$\\
    &PEX&436/432&2.15$_{-0.32}^{+u}$&98$_{-46}^{+450}$&3.1$_{-1.4}^{+3.3}$&-&1.7$_{-0.5}^{+0.1}$&0.91$_{-0.04}^{+0.03}$&2.6$_{-0.2}^{+0.1}$&5.7$_{-0.6}^{+0.5}$\\
    &CP&431/432&1.95$_{-0.07}^{+0.13}$&-&-&20$_{-3}^{+23}$&2$^f$&0.90$\pm$0.03&2.6$\pm$0.1&5.0$_{-0.3}^{+0.5}$\\    
        \hline
       
    PG 1247+267
    &powerlw&366/345&2.13$_{-0.04}^{+0.03}$&-&-&-&-&2.58$\pm$0.06&2.5$_{-0.1}^{+0.2}$&9.4$_{-0.2}^{+0.3}$\\  
    &cutoff&{366/344}&{2.12$_{-0.04}^{+0.02}$}&{$>$220}&-&-&-&2.58$\pm$0.06&2.5$_{-0.1}^{+0.2}$&9.4$_{-0.2}^{+0.3}$\\   
    &PEX&336/343&2.29$_{-0.08}^{+0.09}$&81$_{-28}^{+83}$&2.1$_{-0.9}^{+1.3}$&-&-&2.50$\pm$0.06&2.8$_{-0.2}^{+0.1}$&8.1$\pm$0.3\\  
    &CP&337/342&2.14$\pm$0.04&-&-&$>$19&2.7$_{-1.4}^{+0.4}$&2.50$\pm$0.06&2.8$_{-0.2}^{+0.1}$&7.6$\pm$0.7\\  
    \hline
    \hline
\end{tabular}
\end{table*}

\section{Black Hole Mass of J1634+7031}\label{sec:opt_spec}
The black hole mass of J1634+7031 was estimated in \citet{Shemmer2006} to be log(M$_{\rm BH}$/M$_\odot$)$\sim$10.6 but with a large uncertainty of $\sim$70\% due to the low-quality of the spectrum. We observed the source with FAST on the 1.5-Meter Tillinghast Telescope \citep{Fabricant1998} at the Fred Lawrence Whipple Observatory (FLWO) on UT 2023 March 18 (PI: Zhao). The spectrum was obtained in a configuration of 300 grating lines mm$^{-1}$ with a full wavelength coverage of 3475--7415~\AA, centered at 5450~\AA\ and a total integration time of 1~hr. 

We measured the full width at half maximum (FWHM) of the broad Mg~II emission line to estimate the virial velocity of the gas in the broad line region (BLR) to measure the virial SMBH mass following \citet{McLure2004}. The continuum luminosity was used to estimate the size of the BLR based on the empirical radius-luminosity relation \citep{Bentz2013}. The best-fit black hole is log(M$_{\rm BH}$/M$_\odot$)$\sim$9.9, which is consistent with \citet{Shemmer2006} within their large uncertainties.
 
\section{SED Fitting} \label{sec:SED}
When preparing the ancillary photometric data for SED fitting, we used X-ray fluxes (corrected for absorption) of the power law component in the 0.5--2~keV and 2--10~keV bands derived from the M PEX model, as X-ray data are also modeled using a cutoff power law model in the {\tt CIGALE} code. The optical, near-IR, mid-IR, and radio data are from the Sloan Digital Sky Survey (SDSS) DR 18, Two Micron All Sky Survey \citep[2MASS,][]{Skrutskie2006}, Wide-field Infrared Survey Explorer \citep[WISE,][]{Cutri2021}, and Very Large Array (VLA) FIRST survey \citep{Becker1995}, respectively. J1634+7031 was not detected in SDSS, so its optical photometry is adopted from Dark Energy Spectroscopic Instrument (DESI) Legacy Imaging Surveys DR 9 \citep{Dey2019}. The photometric data are reported in Table~\ref{Tab:multi}. 

The parameters in {\tt CIGALE} are set up following a standard pipeline \citep[e.g.,][]{Yang2022}. The model includes a stellar emission module, a Galactic dust emission module, and an AGN module, which dominates the SED of the sources in our sample as they are luminous type 1 AGN. In the stellar emission component, we adopt a delayed star formation history (SFH), {\tt sfhdelayed} with the parameter choices of {\tt tau\_main} = 1, 2 Gyr and {\tt age\_main} = 0.5, 1, 2 Gyr, and a single stellar population (SSP) assuming the \citet{Bruzual2003} model, {\tt bc03} with {\tt metallicity} = 0.004, 0.02. The nebular templates are based on \citet{Inoue2011} with {\tt logU} = --2, --3 and {\tt zgas} = 0.005, 0.02. We adopt the Galactic dust attenuation (GDA) module, {\tt dustatt\_calzleit} \citep{Calzetti2000,Leitherer2002} with {\tt E\_BV\_young} = 0.1, 0.2 and {\tt E\_BVs\_old\_factor} = 0.44, and the dust emission {\tt dl2014} module \citep{Draine2014} with {\tt alpha} = 2, 3 and {\tt fracAGN} = 0. The AGN emission is modeled by {\tt SKIRTOR} \citep{Stalevski2012,Stalevski2016} with {\tt t} = 7, {\tt q} = 1.0, {\tt oa} = 40$^\circ$, {\tt R} = 20, {\tt i} = 20 (to be consistent with our X-ray spectral fitting), {\tt fracAGN} = 0.9, 0.99, and {\tt EBV} = 0., 0.03, 0.05, 0.1. The X-ray luminosity is connected to the accretion disk 2500~\AA\ luminosity through $\alpha_{\rm OX}$ with {\tt alpha\_ox} = --1.9, --1.8, --1.7, --1.6, --1.5, --1.4, --1.3, --1.2, and {\tt max\_dev\_alpha\_ox}. The radio loudness is set as {\tt R\_agn} = 0.1, 0.5, 1, 5, 10, 20, 30. All photometry have been given an additional 10\% uncertainty, following \citet{Yang2020}. All SED fittings have been visually checked, and their SED shapes have been well modeled with an acceptable reduced $\chi^2$.

\begingroup
\renewcommand*{\arraystretch}{1.3}
\begin{table*}
\begin{center}
\tiny
\caption{Source multiwavelength photometry data. We did not perform SED fitting on lensed sources.}\label{Tab:multi}
  \begin{tabular}{cccccccccccccc}
       \hline
       \hline
    Target&$u$&$g$&$r$&$i$&$z$&$J$&$H$&K$_{\rm s}$&$W1$&$W2$&$W3$&$W4$&1.4~GHz\\ 
    \hline
    Unit&$\mu$Jy&$\mu$Jy&$\mu$Jy&$\mu$Jy&$\mu$Jy&mJy&mJy&mJy&mJy&mJy&mJy&mJy&mJy\\
    \hline
    J0948+4323&213$\pm$3&214$\pm$1&198$\pm$1&308$\pm$2&283$\pm$6&-&-&-&0.38$\pm$0.01&0.44$\pm$0.02&1.8$\pm$0.2&4$\pm$1&3.1$\pm$0.2\\
    J1225+2235&955$\pm$5&994$\pm$3&1193$\pm$4&1485$\pm$6&1779$\pm$10&1.75$\pm$0.04&1.78$\pm$0.07&2.29$\pm$0.07&2.00$\pm$0.04&2.89$\pm$0.06&9.8$\pm$0.2&19$\pm$1&5.7$\pm$0.3\\
    J1259+3423&747$\pm$5&881$\pm$3&1116$\pm$4&1138$\pm$5&1141$\pm$8&0.86$\pm$0.04&1.17$\pm$0.06&0.75$\pm$0.07&1.05$\pm$0.02&1.60$\pm$0.04&4.1$\pm$0.2&10$\pm$1&12.3$\pm$0.6\\
    J1634+7031&-&3429$\pm$16&4932$\pm$21&-&5573$\pm$13&6.96$\pm$0.05&8.91$\pm$0.07&6.99$\pm$0.09&10.3$\pm$0.2&19.6$\pm$0.4&54.7$\pm$0.8&116$\pm$3&0.9$\pm$0.4\\
    J1458+5254&254$\pm$3&315$\pm$2&339$\pm$2&374$\pm$2&336$\pm$5&0.35$\pm$0.04&$<$0.36&$<$0.40&0.31$\pm$0.01&0.54$\pm$0.02&2.2$\pm$0.1&4.9$\pm$0.8&2.0$\pm$0.2\\
    J1233+1612&15$\pm$1&22$\pm$1&55$\pm$1&77$\pm$1&97$\pm$3&0.19$\pm$0.04&0.37$\pm$0.06&0.34$\pm$0.06&0.41$\pm$0.01&0.76$\pm$0.02&2.5$\pm$0.2&8$\pm$1&$<$1\\
     J1454+0324&149$\pm$2&203$\pm$1&238$\pm$2&240$\pm$2&256$\pm$4&-&-&-&0.18$\pm$0.01&0.25$\pm$0.01&1.4$\pm$0.1&2.7$\pm$0.8&4.9$\pm$0.1\\
     J1614+4704&959$\pm$5&927$\pm$3&1116$\pm$4&1651$\pm$6&1686$\pm$9&1.77$\pm$0.07&1.89$\pm$0.08&2.2$\pm$0.1&2.96$\pm$0.07&4.5$\pm$0.08&14.7$\pm$0.3&29$\pm$1&7.3$\pm$0.1\\
     J1247+267&2110$\pm$1&2037$\pm$1&2165$\pm$1&2570$\pm$1&2930$\pm$1&2.93$\pm$0.05&2.99$\pm$0.08&3.56$\pm$0.07&3.19$\pm$0.07&4.57$\pm$0.09&16.8$\pm$0.3&31$\pm$2&$<$1\\
           \hline
              \hline
\end{tabular}
\end{center}
\end{table*}
\endgroup

\begin{figure*}[t] 
\centering
\includegraphics[width=.9\textwidth]{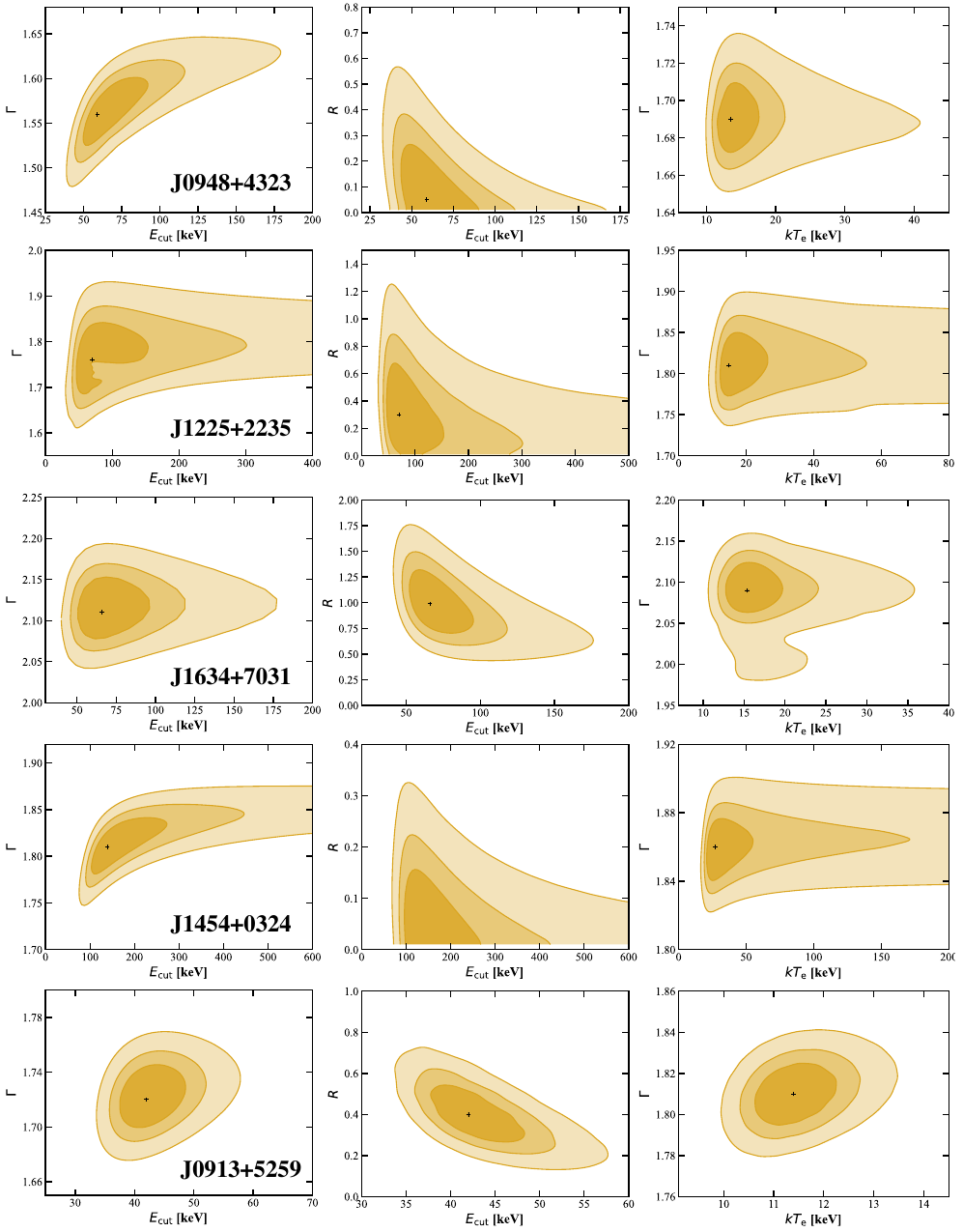}
\caption{Confidence contours, at 68\%, 90\%, and 99\% confidence level, of the parameters $\Gamma$ vs. $E_{\rm cut}$ (left), $R$ vs. $E_{\rm cut}$ (middle) for PEX model and $\Gamma$ vs. $kT_{\rm e}$ (right) for CP model of the PACHA targets.}
\label{fig:contour1}
\end{figure*}  

\begin{figure*}[t] 
\centering
\includegraphics[width=.9\textwidth]{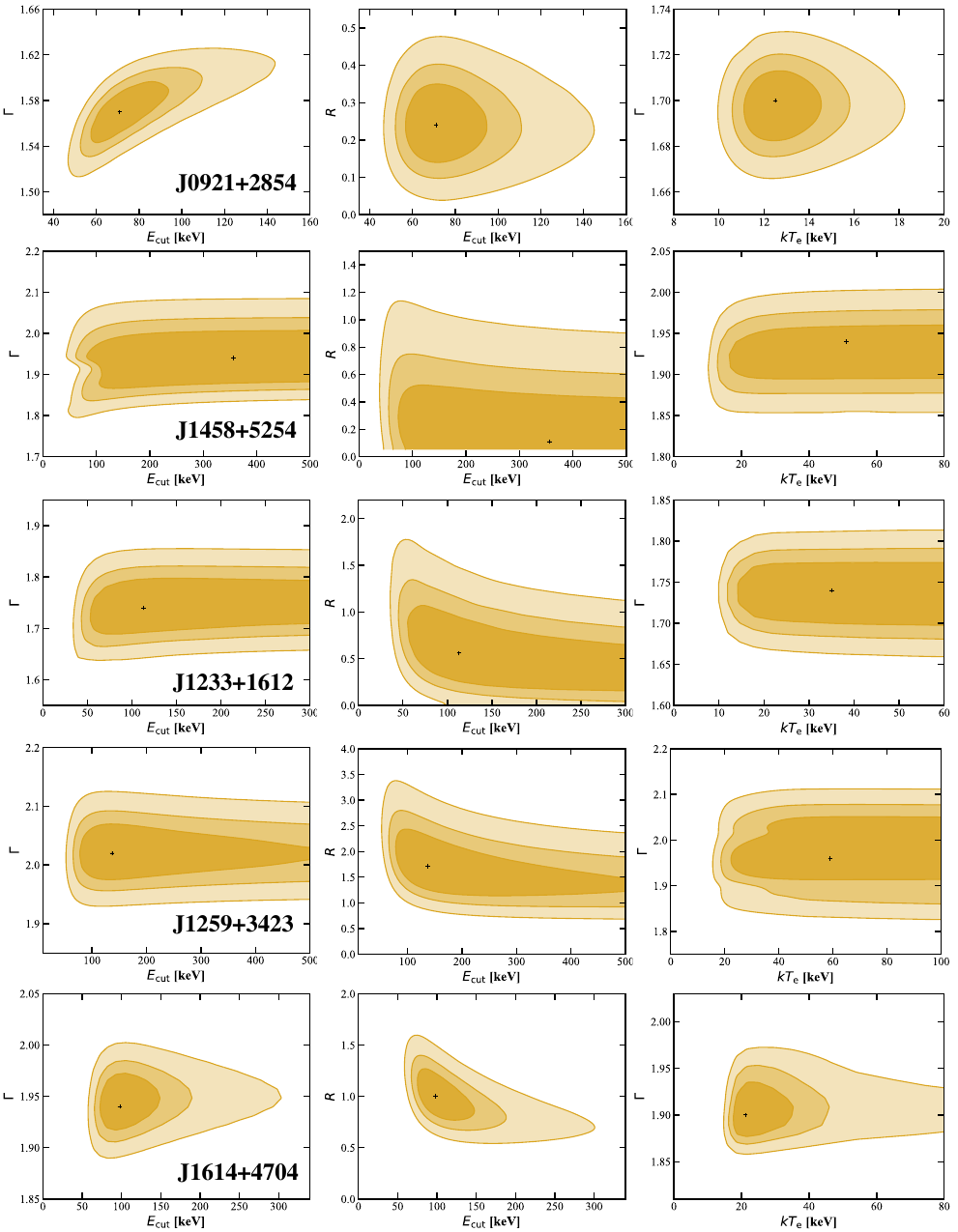}
\caption{Figure~\ref{fig:contour1} continued.}
\label{fig:contour2}
\end{figure*}  

\begin{figure*}[t] 
\centering
\includegraphics[width=.9\textwidth]{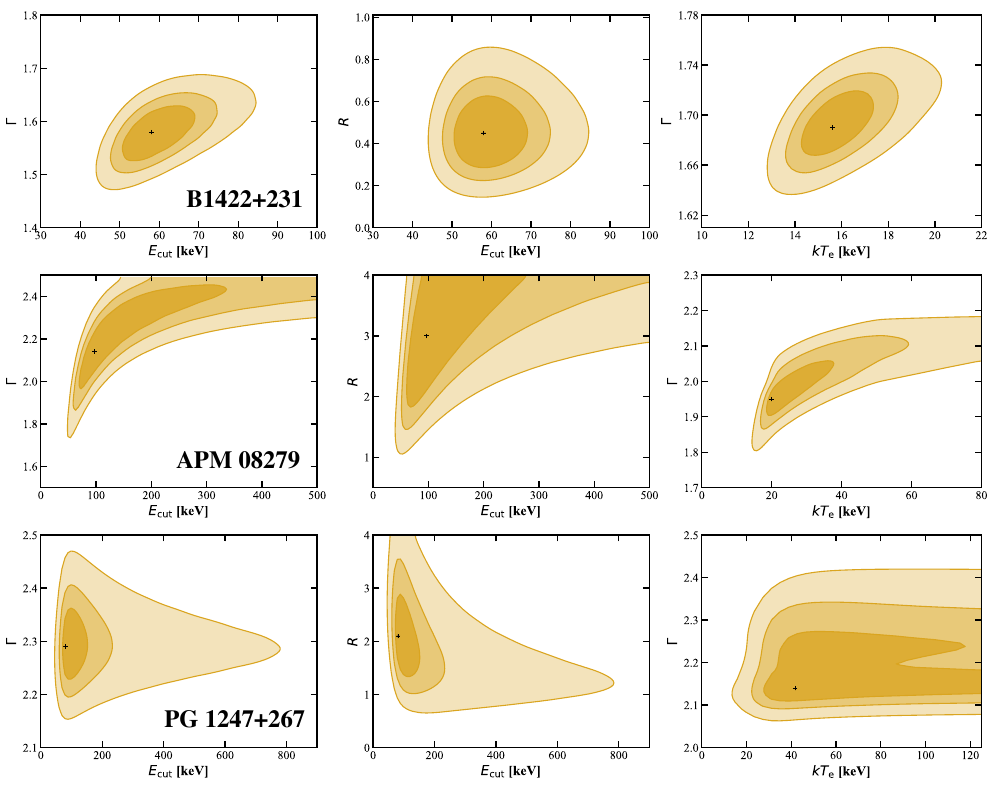}
\caption{Figure~\ref{fig:contour1} continued.}
\label{fig:contour3}
\end{figure*}  

\bibliography{referencezxr}{}
\bibliographystyle{aasjournal}

\end{document}